\documentclass[epj]{svjour}
\usepackage{graphics}
\pdfoutput=1 

\usepackage{color}
\definecolor{brown}{cmyk}{0, 0.8, 1, 0.6}
\definecolor{orange}{rgb}{1,0.5,0}

\usepackage{amsmath}
\usepackage{amssymb}
\usepackage{latexsym}
\usepackage{color}
\usepackage{bm}
\usepackage{multirow}
\usepackage{cancel}
\usepackage{hyperref}

\usepackage{slashed}
\usepackage{comment}
\usepackage{soul}
\setstcolor{red}
\allowdisplaybreaks

\def\beq{\begin{eqnarray}}
\def\eeq{\end{eqnarray}}
\def\beq{\begin{equation}}
\def\eeq{\end{equation}}
\def\be{\begin{eqnarray}}
\def\ed{\end{eqnarray}}

\def\gaa{\mathrel{\raise.3ex\hbox{$>$\kern-.75em\lower1ex\hbox{$\sim$}}}}
\def\la{\mathrel{\raise.3ex\hbox{$<$\kern-.75em\lower1ex\hbox{$\sim$}}}}
\newcommand{\ba}{\begin{array}}
\newcommand{\ea}{\end{array}}
\newcommand{\besub}{\begin{subequations}}
\newcommand{\eesub}{\end{subequations}}
\newcommand{\ee}{\end{equation}}
\newcommand{\bea}{\begin{eqnarray}}
\newcommand{\eea}{\end{eqnarray}}

\begin{document}
\title{Associated production of SM Higgs with a photon in type-II seesaw models at the ILC}
\author{L.~Rahili \inst{1} \and A.~Arhrib\inst{2,3} \and R.~Benbrik \inst{4} 
}                     
%
%
\institute{
EPTHE, Faculty of Science, Ibn Zohr University, P.O.B. 8106 Agadir, Morocco
\and Faculty of Science and Technology, Abdelmalek Essaadi University, P.O.B. 416, Tangier, Morocco 
\and Physics Division, National Center for Theoretical Sciences, Hsinchu, Taiwan 300 
\and MSISM Team, Polydisciplinary Faculty of Safi, Sidi Bouzid, P.O.B. 4162,  Safi, Morocco.
}
\date{Received: date / Revised version: date}
%
\abstract{
We consider the production of a single Standard  Model (SM) Higgs boson ($h^0$) in association with a photon  at the future  Linear collider (LC) in the context of a type-II seesaw model. The type-II seesaw model extends the SM particle content by adding one triple Higgs where its Higgs coupling is a key parameter of the scalar potential triggering the electroweak symmetry-breaking mechanism in the SM. It is a well motivated model since it provides neutrino masses and mixing. The LC presents a particularly interesting possibility to probe Higgs couplings due to the clean beams in the initial state. We study the one loop processes $e^+e^- \to h^0\gamma$ and $e^-\gamma \to e^- h^0$, we show that it is possible to correlate the total cross section predictions with the couplings $h^0\gamma\gamma$ and $h^0\gamma Z$ which are sensitive to the presence of singly and doubly charged Higgs. For parameter points allowed by the current experimental constraints, our numerical results show that the effect of $H^\pm$ and $H^{\pm\pm}$ can be as large as $\pm$ 25\% with respect to the SM predictions. For both processes, we also present differential cross section for different center of mass energy.
} 
\maketitle

\section{Introduction}  
With the discovery of the Higgs-like boson at the Large Hadron Collider (LHC) \cite{atlasdiscovery,cmsdiscovery}, the Standard Model (SM) was finally completed. However, this discovery didn't mark the end of Higgs boson exploration, particularly with regard to mass-generation mechanism for both fermions and gauge bosons. For this purpose, ATLAS and CMS have performed several Higgs coupling measurements, such as $h^0 \to W^+W^-$, $h^0\to ZZ$, 
$h^0\to \gamma \gamma$, $h^0\to b\bar{b}$ and $h^0\to \tau^+ \tau^-$ with a certain precision  which  will be improved during the future run of the LHC. Note that all those  measurements are well consistent with SM predictions. Therefore, any SM extensions should agree with the above measurements.

However, neutrinos are perhaps the most puzzling among all known particles regarding its masses unexplained within the SM \cite{Ellis:2002wba}. Moreover, revealing the true neutrino identity could help answer long-standing questions, like whether neutrino and antineutrino are the same particle, and it could even shade more light on the unification of the interactions that exist in nature. One simple way would be to generate a Dirac neutrino mass in a similar manner that the upper quarks get its own, through the Higgs mechanism by introducing right handed neutrinos, i.e. $m_D\,\bar{\nu_L}\nu_R$. Since neutrino masses are at the sub $\sim eV$ scale, this means that the Yukawa couplings have to be unnaturally small, of the order 10$^{-12}$.

On the other side, if neutrinos happen to coincide with anti-neutrinos, then a non-vanishing dimension five operator $\frac{1}{\Lambda}\mathcal{L}_5$ \cite{Weinberg:1979sa,Wilczek:1979hc}, constructed from the neutrino and Higgs fields, gives rise to a light Majorana neutrino mass, and leads to leptonic number violation (global symmetries are not sacred). In this case, the large Majorana mass scale $\Lambda$ that turns out to be quite close to the Grand Unification scale, ends up suppressing the neutrino masses via Seesaw mechanism as $v_{EW}^2/\Lambda$. 

There are three types of Seesaw mechanisms that provide a
comprehensive understanding of the observed sizes of neutrino masses by exchange of 3 different types of heavy particles, known as $i)$ Type-I seesaw where a gauge singlet chiral fermion $S=(1,1,0)$ is introduced to the SM sector \cite{Minkowski:1977sc,Mohapatra:1979ia,Yanagida:1979as,GellMann:1980vs,Schechter:1980gr,Glashow:1979nm,Babu:1993qv,Antusch:2001vn}, $ii)$Type-II seesaw in which a $SU(2)_L$ weak triplet boson $\Delta=(1,3,2)$ is added to the SM Higgs sector~\cite{Magg:1980ut,Cheng:1980qt,Lazarides:1980nt,Mohapatra:1980yp} and finally $iii)$ the addition of a $SU(2)_L$ triplet fermion $T=(1,3,0)$~\cite{Foot:1988aq} provides the conventional Type-III Seesaw.

So among the previous categories, Seesaw Type-II, also known as the Higgs Triplet Model (HTM), remains the viable option that is made up of the SM Higgs doublet and a triplet of hypercharge of $Y = +2$. In this context, once the neutral component of the triplet $\Delta$ develops its own vacuum expectation value, the neutrinos acquire Majorana mass. In this regards, the most significant attribute of HTM is its minimality regarding the very simple representations of $SU(3) \times\,SU(2)\,\times\,U(1)$, for both fermions and Higgs sectors.

A close look at the HTM Higgs sector reveals a variety in its spectrum which includes: 2 CP-even Higgs $h^0$ and $H^0$, one CP-odd $A^0$, pair of charged Higgs $H^\pm$ and a pair of doubly charged Higgs $H^{\pm \pm}$. For more details about HTM spectrum as well as the theoretical constraints, we refer the interested readers to Refs \cite{Arhrib:2011uy,Dev:2013ff}. Moreover,  a number of detailed phenomenological researchs have already been performed at the LHC \cite{Perez:2008ha,Melfo:2011nx,delAguila:2008cj,Chakrabarti:1998qy,Aoki:2011pz,Chun:2013vma,Banerjee:2013hxa}. One attractive feature of this model is the presence of the doubly-charged Higgs boson, and its distinguished decay modes which strongly depend on the size of the triplet {\it vev}.

At the LHC, ATLAS and CMS already performed the measurement of several Higgs couplings to SM particles with an uncertainty of about 10-20$\%$. These measurements will be further improved by the High 
Luminosity option of the LHC (HL-LHC) which brings uncertainties down to 2-5$\%$ \cite{Gianotti:2000tz}.
Moreover, at the $e^+e^-$ Linear Collider (LC), which would act like a Higgs factory, 
the uncertainties on the Higgs couplings would be much smaller, reaching 0.6-1.3\% for some light fermionic decay channels \cite{Moortgat-Picka:2015yla}. It is well known that the precise measurement programs of Higgs properties at the LC and LHC are complementary to each other in many aspects.
In addition, the loop-mediated process $h^0\to \gamma\gamma$ which was a discovery mode for 
the 125 GeV Higgs-like and is now quite accurately measured.  While the other related one-loop decay $h^0\to \gamma Z$, which is also loop-mediated, is still missing and may show up in the future LHC run when more data is accumulated.
At the LC, the one loop mediated process $e^+e^-\to \gamma h^0$ if measured could also shed some light on $h^0\gamma \gamma$ and $h^0\gamma Z$ couplings. Such process has been investigated in the SM \cite{Barroso:1985et,Abbasabadi:1995rc}, and also in many Beyond SM (BSM) scenarios, like SUSY~\cite{Djouadi:1996ws,Demirci:2019ush},  extended Higgs sector \cite{Akeroyd:1999gu,Kanemura:2018esc}, and in the Inert Higgs Model  \cite{Arhrib:2015wrb}. 

In this work, we examine the effect of a singly and a doubly charged Higgs bosons in 
the associated production of the SM Higgs with a photon $e^+e^- \to h^0 \gamma$ at the LC and also $e^-\gamma \to h^0 e^-$ which would take place if the  $e^-\gamma$ collisions are available at the LC.  In Ref\cite{Kanemura:2018esc}, the authors study $e^+e^- \to \gamma h^0$ in the framework of Inert Triplet Model  with an exact $Z_2$ symmetry under which 
the triplet scalar is odd while all the other SM particles are even which guaranty that the model have a dark matter candidate. Because of $Z_2$ symmetry, the SM Higgs comes only from the doublet, 
the couplings  $h^0H^\pm H^\mp$ and $h^0H^{\pm \pm}H^{\mp\mp}$ originate only from triplet interaction with SM doublet while in our case there is extra terms that come 
from the mixing between the doublet and triplet components.
In addition, in our study, we will also address the process  $e^- \gamma \to h^0 e^-$ 
which is not covered in Ref\cite{Kanemura:2018esc}.

Study of these processes can be used to shed some light on the couplings $h^0\to \gamma \gamma$ and $h^0\to \gamma Z$ and their correlation. Clearly, such loop-mediated processes are sensitive to the $h^0\gamma\gamma$ and $h^0\gamma Z$ one loop couplings, and could also be used to disentangle between various BSM models.
Therefore,  $e^+e^- \to h^0 \gamma$ and $e^- \gamma \to h^0 e^-$  would be sensitive to the 
doubly charged Higgs  which contribute to $h^0\gamma \gamma$ and $h^0\gamma Z$ couplings.
Moreover, the process $e^+e^- \to h^0 \gamma$ enjoys a clean final state with a photon 
and also the handle offered by the SM-like Higgs mass reconstruction at 125 GeV, which is now possible after discovery.
We will study the correlation of the diphoton signal strength with the total cross sections of 
$e^+e^- \to h^0 \gamma$ and $e^- \gamma \to h^0 e^-$ and also present some differential cross sections. 

The remainder of this paper is organized as follows: we briefly review the basics of the Triplet Higgs Model in Sec.~\ref{model}. In Sec.~\ref{expcons}, we discuss existing experimental constraints. In the subsequent subsections \ref{modsiglow} and \ref{modsighigh}, we consider in more depth the production cross-sections of $e^+e^-\to h^0\gamma$ and 
$e^- \gamma \to h^0 e^-$ at the $e^{+}e^{-}$ collider and its $e^-\gamma$ option.  We also correlate the production cross section to $h^0\to \gamma\gamma$ and $h^0\to \gamma Z$ decays. We then present our findings in Sec.~\ref{conclu}.
\section{Model framework}
\label{model}
In addition to the SM Higgs field $\Phi\sim(1,2,1/2)$, the HTM contains an additional  $SU(2)_L$ triplet Higgs field $\Delta\sim(1,3,2)$ \cite{Magg:1980ut,Cheng:1980qt,Lazarides:1980nt,Mohapatra:1980yp}. Both multiplets are represented by 
\begin{eqnarray}
\Phi =
\left(
\begin{array}{c}
\phi^+ \\
\phi^0
\end{array}
\right),
\quad
\text{and}
\quad
\Delta =
\left(
\begin{array}{cc}
\frac{\delta^+}{\sqrt{2}} & \delta^{++} \\
\delta^0 & -\frac{\delta^+}{\sqrt{2}} \\
\end{array}
\right)
\label{multiplet}
\end{eqnarray}
After the EWSB, the neutral components $\phi^0$ and $\delta^0$ 
develop their {\it vev}'s respectively that read as
$$
\phi^0 =\frac{1}{\sqrt{2}}(v_\Phi+\eta_\Phi+i\chi_\Phi)  \quad
\text{and}
\quad
\delta^0=\frac{1}{\sqrt{2}}(v_\Delta+\eta_\Delta+i\chi_\Delta)
$$
It's worth mentioning that the gauge bosons get their masses both from the doublet 
and the triplet and the electroweak scale is fixed from W mass by $v^2=v^2_{\Phi}+2\,v^2_{\Delta}=(246 \, \, \rm{GeV})^2$. 

On the other side, the kinetic and gauge-interaction of the new field $\Delta$ are encoded in the lagrangian term that has the following form
\begin{eqnarray}
\mathcal{L}_{\rm{kin}}( \Delta)&=&\rm{Tr}[(D_\mu \Delta)^\dagger (D^\mu \Delta)], 
\label{kinetic}
\end{eqnarray}
where the covariant derivative is given by $D_\mu \Delta=\partial_\mu \Delta+i\frac{g}{2}[\tau^aW_\mu^a,\Delta]+ig'B_\mu\Delta$.
The Yukawa interactions of $\Delta$ with the lepton fields are 
\begin{eqnarray}
\mathcal{L}_Y(\Phi, \Delta)&=& Y_{\Delta}\bar{L}_L^{c}i\tau_2\Delta L_L+\rm{h.c.}.~~~~ 
\label{yukawa}
\end{eqnarray}
In the above, $c$ denotes the charge conjugation, while $L$ is the $SU(2)_{L}$ doublets of left-handed leptons. Once $\Delta$ is assigned to carry a non-zero lepton number, $L=2$, the cubic $\mu$ parameter violates explicitly the lepton number at the Lagrangian level. It's worth mentioning here that the light neutrino mass originate from the above Yukawa lagrangian and 
is proportional to the triplet {\it vev}, 
$m_\nu \sim v_{\Delta} Y_\Delta/\sqrt{2}$, where $Y_\Delta$ is the Yukawa coupling.

The scalar potential of the Higgs fields $\Phi$ and $\Delta$ is  
\begin{eqnarray}
V(\Phi,\Delta)&=&\mu_\Phi^2\Phi^\dagger\Phi+\mu^2_{\Delta}\rm{Tr}(\Delta^\dagger\Delta)+\left(\mu \Phi^Ti\tau_2\Delta^\dagger \Phi+\rm{h.c.}\right)\nonumber\\
&+&\frac{\lambda}{4}(\Phi^\dagger\Phi)^2 + \lambda_1(\Phi^\dagger\Phi)\rm{Tr}(\Delta^\dagger\Delta)+\lambda_2\left[\rm{Tr}(\Delta^\dagger\Delta)\right]^2\nonumber\\
&+&\lambda_3\rm{Tr}[(\Delta^\dagger\Delta)^2]
+\lambda_4\Phi^\dagger\Delta\Delta^\dagger\Phi,~~~~
\label{eqn:scalpt}
\end{eqnarray}
where $\mu_{\Phi}$ and $\mu_\Delta$ are real parameters with dimension of mass, 
$\mu$ is a mass term that controls trilinear couplings between $\Phi$ and $\Delta$ and $\lambda_{1-4}$ are dimensionless quartic Higgs couplings.

By computing the mass matrices from Eq.(\ref{eqn:scalpt}) taking into account the two minimization conditions, seven physical Higgs states in the mass basis arise and could be written in the weak eigenstate basis. Indeed, in addition to the doubly-charged Higgs bosons, $H^{\pm \pm}$, the HTM predicts a pair of charged Higgs $H^\pm$, that appear together with the charged Goldstone $G^{\pm}$ after a unitary rotation $R_{\beta^{'}}=\{\{c_{\beta^{'}},-s_{\beta^{'}}\},\{s_{\beta^{'}},c_{\beta^{'}}\}\}$ with $\beta'$ angle between the non-physical fields $\phi^{\pm}$ and $\delta^{\pm}$, with $s_x\,(c_x)$ stands for $\sin x\,(\cos x)$.

Similarly, two unitary rotations, $R_{\alpha}$ and $R_{\beta}$ in the planes $(\eta_\Phi,\eta_\Delta)$ and $(\chi_\Phi,\chi_\Delta)$, give rise respectively to two CP-even neutral scalars $(h^0,H^0)$ and two CP-odd neutral pseudo-scalars $(G^0,A^0)$.

The physical masses of the doubly and singly charged Higgs boson $H^{\pm \pm}$ and $H^{\pm}$  
can be written as
\begin{eqnarray}
m_{H^{++}}^2&=&m_\Delta^2-v_\Delta^2\lambda_3-\frac{\lambda_4}{2}v_{\Phi}^2,\quad m_\Delta^2 = \frac{\mu v_\Phi^2}{\sqrt{2}v_\Delta}\label{eq:mhpp}\nonumber\\
m_{H^+}^2&=&\left(m_\Delta^2-\frac{\lambda_4}{4}v_{\Phi}^2\right)\left(1+\frac{2v_\Delta^2}{v_{\Phi}^2}\right).\label{eq:mhp}
\end{eqnarray}
The CP-even and CP-odd neutral Higgs bosons $h^0$, and $H^0$  have the physical masses
\begin{eqnarray}
m^2_{h^0}&=&\mathcal{A}_{H}^2 c_\alpha^2+\mathcal{C}_{H}^2 s_\alpha^2-\mathcal{B}_{H}^2 s_{2\alpha}, \label{mh}\nonumber\\
m^2_{H^0}&=&\mathcal{A}_{H}^2 s_\alpha^2+\mathcal{C}_{H}^2 c_\alpha^2+\mathcal{B}_{H}^2 s_{2\alpha}.\label{mH}
\end{eqnarray}
In the above $\mathcal{A}_{H}$, $\mathcal{B}_{H}$ and $\mathcal{C}_{H}$ are the entries of the CP-even mass matrix, given by, 
\begin{eqnarray}
&&\mathcal{A}_{H}^2=\frac{v_{\Phi}^2\lambda}{2},\nonumber\\
&&\mathcal{B}_{H}^2=-\frac{2v_\Delta}{v_{\Phi}}m_\Delta^2+v_{\Phi} v_\Delta \lambda_{14}^+,\nonumber\\
&&\mathcal{C}_{H}^2=m_\Delta^2+2v_\Delta^2 \lambda_{23}^+.
\end{eqnarray}
where $\lambda_{ij}^+=\lambda_i+\lambda_j$, and $s_{2x}$ stands for $\sin 2x$.

Regarding the CP-odd Higgs field, the $A^0$ has the mass term given by
\begin{eqnarray}
m_{A^0}^2 &= &m_\Delta^2\left(1+\frac{4v_\Delta^2}{v_{\Phi}^2}\right) \label{mA}.
\end{eqnarray}
In addition, it should be noted that the previously mentioned mixing angles can be determined in terms of the multiplets {\it vev}'s and the dimensionless couplings as follows,
\begin{eqnarray}
&&t_\beta = \sqrt{2}\,t_{\beta^{'}} = 2 v_\Delta/v_\Phi\label{eq:91} \label{eq:tb-tbp}\\
&&t_{2\alpha} = \frac{2\mathcal{B}_{H}^2}{\mathcal{A}_{H}^2-\mathcal{C}_{H}^2}\quad ({\rm with}\,\, t_x=\tan x), \label{eq:ta}
\label{eq:92}
\end{eqnarray}
Furthermore, we consider throughout our analysis a different hierarchy between $m_{H^{\pm \pm}}$ and $ m_{H^{\pm}}$, which mainly depends on $\lambda_4$ sign (e.g. for positive $\lambda_4$, the ${H^{\pm \pm}}$ is lighter than ${H^{\pm }}
$), leading to a splitting that can be expressed by (assuming $v_{\Delta} \ll v_{\Phi}$)
\begin{equation}
\Delta m^2=m^2_{H^{\pm}}-m^2_{H^{\pm \pm}} \sim \frac{\lambda_4}{4} v^2_{\Phi}+\mathcal{O}(v^2_{\Delta}).
\label{diffchdmass}
\end{equation}

It is worth noting that it is possible to invert these masses to write the quartic couplings $\lambda$ and $\lambda_i$'s
and $\mu ({\rm or}\,M^2_{\Delta})$ in terms of the 5 physical scalar masses and the mixing angles as done in \cite{Arhrib:2011uy}. Therefore, in our numerical investigation, to fully describe the scalar sector of HTM, we will take the lightest $h^0$ as SM-like, and yet use the following set of parameters 
\beq
{\cal P} = \{\lambda, \lambda_1, \lambda_2, \lambda_3, \lambda_4 , \mu , v_{\Delta} \}
\label{eq:input_par_2}
\eeq
  
The following section describes the various direct experimental constraints on the doubly-charged Higgs boson mass and triplet {\it vev}.
\section{Theoretical and Experimental Constraints}
\label{expcons}
The parameter space of HTM discussed above is subjected to both
theoretical and experimental constraints as we will describe briefly here.
\begin{itemize}
\item [$\circ$] {Perturbativity and Unitarity:} 
The perturbativity translates into the requirement that all quartic couplings of the scalar potential in
Eq.~(\ref{eqn:scalpt}) obey $|\lambda_i| \le 8 \pi $. 
Tree-level unitarity can also be imposed by considering
a variety of scattering processes: scalar-scalar scattering,
gauge boson-gauge boson scattering and scalar-gauge boson scattering.
We impose these unitarity constraints as derived in~\cite{Arhrib:2011uy}.
\item [$\circ$]  {Vacuum Stability:}
By requiring the HTM to satisfy vacuum stability, we order to maintain the scalar potential $V$ bounded from below,
the following constraints on the HTM parameters must be met~\cite{Arhrib:2011uy,bonilla2015}
\begin{eqnarray}
&& \lambda \geq 0 \;\;{\rm \&}  \;\; \lambda_{23}^+ \geq 0  \;\;{\rm \&}  \;\;\tilde\lambda_{23}^+ \geq 0 \label{eq:bound1} \\
&& {\rm \&} \;\;\lambda_1+ \sqrt{\lambda \lambda_{23}^+} \geq 0 \;\;{\rm \&}\;\;\lambda_{14}^++ \sqrt{\lambda \lambda_{23}^+} \geq 0  \label{eq:bound2} \\
&& {\rm \&} \;\; 2 \tilde{\lambda}_{14}+\sqrt{(2\lambda \lambda_3-\lambda_4^2) (2\frac{\lambda_2}{\lambda_3} + 1)} \geq 0 \;\; {\rm or} \nonumber\\
&&\hspace{2cm} \lambda_3 \sqrt{\lambda} \le |\lambda_4| \sqrt{\lambda_{23}^+} \label{eq:bound3}
\end{eqnarray}
where $\tilde\lambda_{ij}^+=\lambda_i+\frac{1}{2}\lambda_j$.
\end{itemize}

Furthermore, collider signatures analysis of doubly charged Higgs production could lead to spectacular signature in some cases.
Depending on the decay modes of $H^{\pm\pm}$ to either leptonic or bosonic final states, one could have the popular same sign leptons  signature. 
To that end, the triplet {\it vev} $v_{\Delta}$  size
might be a decisive point. For smaller triplet {\it vev}, the $H^{\pm \pm}$ predominantly decays into the same-sign leptonic states $H^{\pm \pm } \to l^{\pm} l^{\pm}$, whereas for  larger $v_{\Delta}$, the gauge boson mode $H^{\pm \pm} \to W^{\pm} W^{\pm}$ becomes dominant \cite{Perez:2008ha,Melfo:2011nx}.  A number of studies have been proposed in order to study prospect for doubly charged Higgs production and decay at the LHC \cite{Perez:2008ha,Melfo:2011nx,delAguila:2008cj,Mitra:2016wpr,Agrawal:2018pci}. \\ 
Below we discuss the existing constraints on  $H^{\pm \pm}$ from collider searches. 
\\
\begin{itemize}
\item LEP-II: LEP-II detector was used to search for doubly charged Higgs through it decay $H^{\pm \pm} \to l^\pm l^\pm$. The limit obtained 
is $m_{H^{\pm \pm}} > 97.3$ GeV \cite{Abdallah:2002qj} at 95$\%$ C.L.
\\
\item LHC: constraints from $H^{\pm\pm}$ pair production and associated production with 
$H^\pm$ at the LHC with 13 TeV set a rigorous constraint on $m_{H^{\pm \pm}}$ for  $v_{\Delta} < 10^{-4}$ GeV through the same-sign dilepton decay $H^{\pm \pm } \to l_i^{\pm } l_j^{\pm}\,(i,j=e,\mu,\tau)$. In addition, the CMS searches also include the associated production $p p \to H^{\pm \pm} H^{\mp}$ followed by $H^{\pm} \to l^{\pm} \nu$. This combined channel of pair-production and associated production gives the stringent constraint $m_{H^{\pm \pm}} > 820$ GeV \cite{CMS-PAS-HIG-16-036} at $95 \%$ C.L for $e$ and  
 $\mu$ flavor. The main constraint comes from ATLAS searches through 
 pair-production, which set a lower limit of $m_{H^{\pm\pm}} > 870$ GeV at $95 \%$ C.L \cite{Aaboud:2017qph}.
As discussed above, if the triplet {\it vev} is larger then $H^{\pm\pm}$ would decay into a pair of gauge bosons \cite{Kanemura:2013vxa,Kanemura:2014goa,Kanemura:2014ipa}, and this would invalidate or, lower the same-sign dileptons limit.
Over the past year, the ATLAS experiment has managed in this regard to set a limit for $H^{\pm\pm}$ mass, 
in such a way that the same-sign W bosons decay mode to be the dominant for doubly-charged Higgs boson. As stipulated by its report, a $H^{\pm\pm}$ boson masses between 200 and 220 GeV are excluded at $95\%$ confidence level \cite{Aaboud:2018qcu}.

Furthermore, still within the LHC, the golden channel that can give more insights for the doubly-charged Higgs boson is the VBF $pp\to W^*W^* \to H^{\pm\pm}\to W^\pm W^\pm$. In the HTM, it is well known that once the triplet {\it vev} gets larger enough, {$v_{\Delta} \ge 10^{-4}$ GeV}, the bosonic decay mode $H^{\pm\pm} \to W^\pm W^\pm$ becomes larger than $BR(H^{\pm\pm} \to l^\pm l^\pm)$. At the LHC, CMS with 8 TeV \cite{Khachatryan:2014sta} and 13 TeV \cite{Sirunyan:2017ret} search for $pp\to W^*W^*\to H^{\pm\pm} \to W^\pm W^\pm+X$ in the framework of Georgi-Machacek (GM) model and set a limit on the doublet-triplet mixing angle $s_H(=\sin\theta_H)=2\sqrt{2}v_\chi/v$ where $v_\chi$ is the triplet {\it vev} in the GM model. Since in the HTM $H^{\pm\pm}W^\pm W^\pm = i g^2 v\,s_{\beta^{'}}/\sqrt{2}$  and $ H^{\pm\pm}W^\pm W^\pm=i g^2 v\,s_{H}/\sqrt{2}$ in the GM, it is clear that the equivalent of $s_H$ in the HTM is $s_{\beta^{'}}$ that can be obtained from Eq.(\ref{eq:tb-tbp}) as: $s_{\beta^{'}}=2v_\Delta/v$. Re-interpretation of GM limit on $s_H$ imply that $s_{\beta^{'}} > 0.18$  which can be translated into a limit on the triplet {\it vev}.  Using Eq.(\ref{eq:tb-tbp}), one can show that $v_\Delta \ll v_\Delta^{max} = v\,{\rm max}[s_{\beta^{'}}]/2 \approx 22.14$ GeV which is still far from the stringent bound obtained from $\delta\rho$ constraint.
\end{itemize}

\noindent
Note that, for extremely small $v_{\Delta}$, the mass of the doubly-charged Higgs  boson is very tightly constrained from pair-production searches. For a larger triplet {\it vev}, this constraint substantially chills out. Indeed, the measured cross-section of the VBF production process probes a quadratically reliance upon the triplet {\it vev} and hence, increases for a very large {\it vev}. However, the range of $v_{\Delta} \sim 10^{-4}-10^{-1}$ GeV cannot be probed at the 13 TeV LHC in VBF channel, due to the smallness 
 of  the production cross-section. Recently, in \cite{Ghosh:2017pxl}, 
the authors have looked for pair-production  of 
 doubly charged Higgs $H^{\pm \pm}$ in the case of large $v_{\Delta}$, and demonstrated that the lighter mass $m_{H^{\pm \pm} } \sim 190$ GeV can be probed at the 14 TeV LHC with high luminosity of 3000 $\rm{fb}^{-1}$. In addition, in Ref.\cite{Kanemura:2014ipa}, by looking to di-lepton decays of $H^{\pm \pm}$ and using LHC 8 TeV run-I data, the limit of  $m_{H^{\pm \pm}} \ge 84 $ GeV have been derived, which is also relevant for large $v_{\Delta}$.

Furthermore, all along our computing, the latest version of {\tt HiggsBounds-5.3.2beta} and 
{\tt HiggsSignals-2.2.3beta} \cite{Bechtle2014:HB} \cite{Bechtle2014:HS} have been used to conveniently test Higgs searches and check the Higgs signal rate constraints in the HTM taking into account various LEP, Tevatron and recent LHC 13 TeV search results.

\section{$e^+e^- (e^- \gamma) \to \gamma h^0 (e^- h^0)$ in type-II seesaw models}
\label{section:3}
\subsection{Processes}
\label{modsiglow}

\begin{figure*}[!h]
\centering
\resizebox{0.7\textwidth}{!}{
\includegraphics{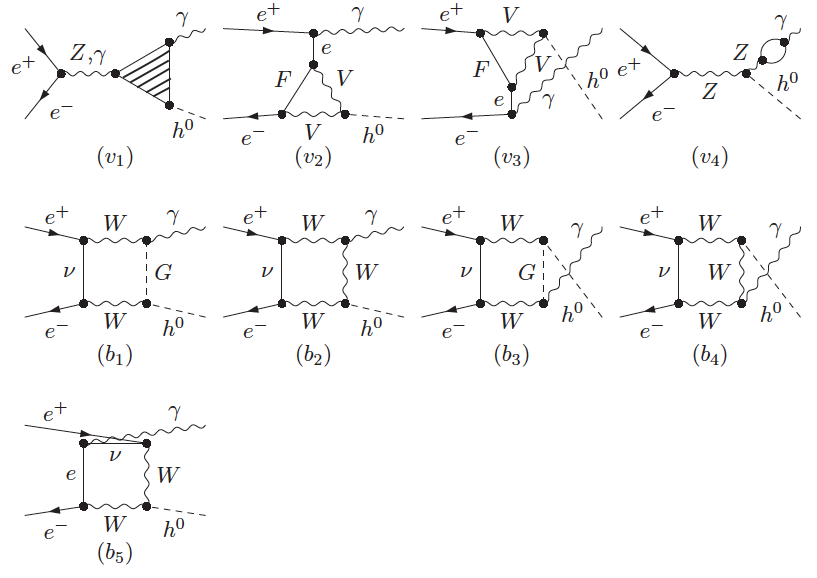}}
\caption{Generic Feynman diagrams involving the various contributions to $e^-\,e^+\,(e^-\,\gamma)\,\to\,\gamma\, h^0\,(e^-\, h^0)$ processes in the \textsc{HTM}. 
In diagrams $v_1$, we depict $h^0\gamma V$, $V=\gamma,Z$ as a generic and the feynman diagrams that contribute are shown in Fig.\ref{fig:diag-h2gaV}. The loop in $v_4$ receives
contributions from all SM particles as well as  singly and doubly charged Higgs.}
\label{fig:diag-ee2gah}
\end{figure*}

In the HTM, at the one-loop level, the processes $e^+e^-\to \gamma h^0$ and
 $e^- \gamma \to e^- h^0$ are mediated by triangle and self-energie as well as box diagrams. Hence, they are sensitive to all charged virtual particles inside the loop. 
We display in Fig.\ref{fig:diag-ee2gah} the generic Feynman diagrams that effectively contribute to 
$e^+e^- \to \gamma h^0$ and $e^- \gamma \to e^- h^0$ processes in the HTM. 

Our calculation is done in Feynman gauge using dimensional regularization 
with the help of {\sl FeynArts} and {\sl FormCalc} packages \cite{FA2}. 
Numerical evaluation of the scalar integrals is done with {\sl LoopTools} \cite{FF}. 
We have summed up all triangle and boxe diagrams in order to maintain gauge invariance in 
the final results. As said before, during this calculation we will neglect the electron mass. 
Since the tree level amplitudes which are suppressed by the electron mass are neglected, 
Feynman diagrams like Fig. \ref{fig:diag-ee2gah}-$v_2$-$v_3$\footnote{$v_2$ and $v_3$ are negligible compare to other diagrams} mediated by an off-shell electron are ultraviolet finite because the corresponding counter-terms for $h^0e^+e^-$ are proportional to electron mass.
We have checked analytically and numerically that the amplitudes are UV finite 
and independent of the renormalization scale which constitutes a good check of the calculation.

\begin{figure*}[!h]
\centering
\resizebox{0.8\textwidth}{!}{
\includegraphics{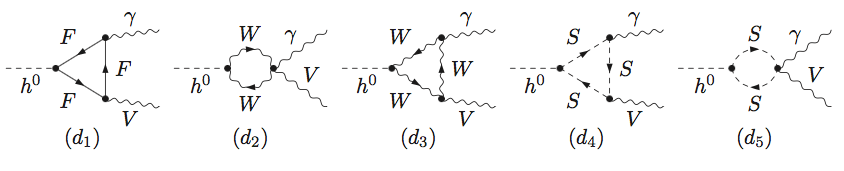}}
\caption{Generic Feynman diagrams for $h^0 \to \gamma\,V$ ($V=\gamma\,{\rm or}\,Z$) process in the \textsc{HTM}. F denote any fermion and S stand for $H^{\pm}$ and $H^{\pm\pm}$.}
\label{fig:diag-h2gaV}
\end{figure*}

In the following, for illustrative purpose and discussions, we introduce the following ratios,

\begin{eqnarray}
&& R_{\gamma h^0}  \equiv  
\frac{\sigma(e^+e^-\to \gamma h^0)}{\sigma_{\rm SM}(e^+e^- \to \gamma H)}\nonumber\\
&& R_{ e h^0}  \equiv  
\frac{\sigma(e^- \gamma \to e^- h^0)}{\sigma_{\rm SM}(e^- \gamma  \to e^- H)}\nonumber
\end{eqnarray}
which are the total cross sections in the HTM normalized to the SM one.  
Note that the one-loop amplitudes for $h^0\to\gamma \gamma, \gamma Z$, Fig. \ref{fig:diag-h2gaV}, 
as well as for the two processes $e^+e^-\to \gamma h^0$ and $e^-\gamma \to e^- h^0$ receive an additional contribution from  
$H^\pm$ and $H^{\pm\pm}$ Higgs bosons.

We also define the signal strengths $R_{\gamma\gamma}$ and $R_{\gamma Z}$ as
\begin{equation}
R_{\gamma V}  \equiv  \frac{\sigma(gg \to h^0)\times 
Br(h^0\to\gamma V)}{\sigma_{SM}(gg \to H)\times Br_{SM}(H\to\gamma V)},
\end{equation}
where $V=\gamma\,{\rm or}\,Z$

These one-loop amplitudes are sensitive to the triple scalar 
couplings $h^0 H^{++}H^{--}$ and $h^0 H^{+}H^{-}$ which are given in the HTM by
\begin{eqnarray}
\bar{\lambda}_{h^0 H^{++}H^{--}} &=&\frac{s_W}{e\,m_W}\big(2\,\lambda_2\,v_\Delta\,s_\alpha+\lambda_1\,v_{\Phi}\,c_\alpha\big)
 \label{eq:redgcalHHpp}\\
\bar{\lambda}_{h^0 H^+H^-}&=&\frac{s_W}{2\,e\,m_W}\Big(\big(4\,v_\Delta\, \lambda_{23}^+ c_{\beta'}^2+2\,v_\Delta\lambda_1\,s_{\beta'}^2\nonumber\\
&-&\sqrt{2}\,\lambda_4\,v_{\Phi}\,c_{\beta'}\,s_{\beta'}\big)\,s_\alpha +\big(\lambda\,v_{\Phi}\,s_{\beta'}^2+2 \tilde\lambda_{14}^+\,v_{\Phi}\,c_{\beta'}^2\nonumber\\
&+&(4\,\mu-\sqrt{2}\,\lambda_4\,v_\Delta)\,c_{\beta'}s_{\beta'}\big)c_\alpha\Big)
\label{eq:redgcalHHp}
\end{eqnarray}
If we neglect the terms which contain triplet {\it vev}, those couplings are completely fixed by the $\lambda_1$ and $\lambda_4$ parameters, and depending on their sign; charged Higgs contributions can enhance or 
suppress $e^+e^-\to \gamma h^0$, $e^-\gamma \to e^- h^0$  and $h^0\to\gamma \gamma$ rates, respectively. We also stress that the doubly charged Higgs contribution in the above one-loop amplitudes will enjoy 
an enhancement by a relative factor four in the amplitudes since $H^{\pm\pm}$ 
has an electric charge of $\pm 2$ units.

\subsection{Results}
\label{modsighigh}

\begin{figure*}[!ht]
\centering
\begin{minipage}{6.1cm}
\begin{center}
\resizebox{1.1\textwidth}{!}{\includegraphics{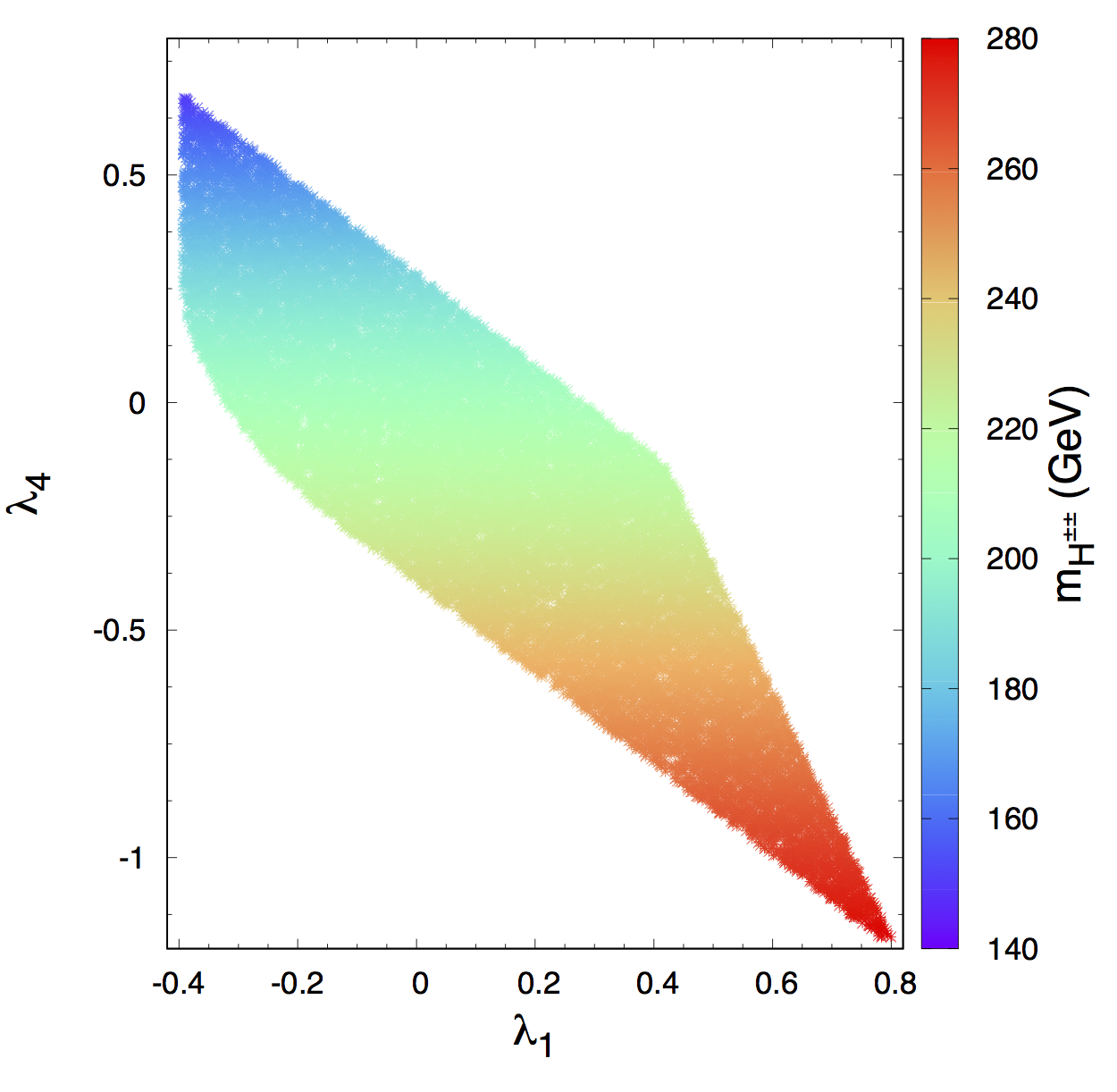}}
\end{center}
\end{minipage}
\hspace{0.6cm}
\begin{minipage}{6.1cm}
\begin{center}
\resizebox{1.1\textwidth}{!}{\includegraphics{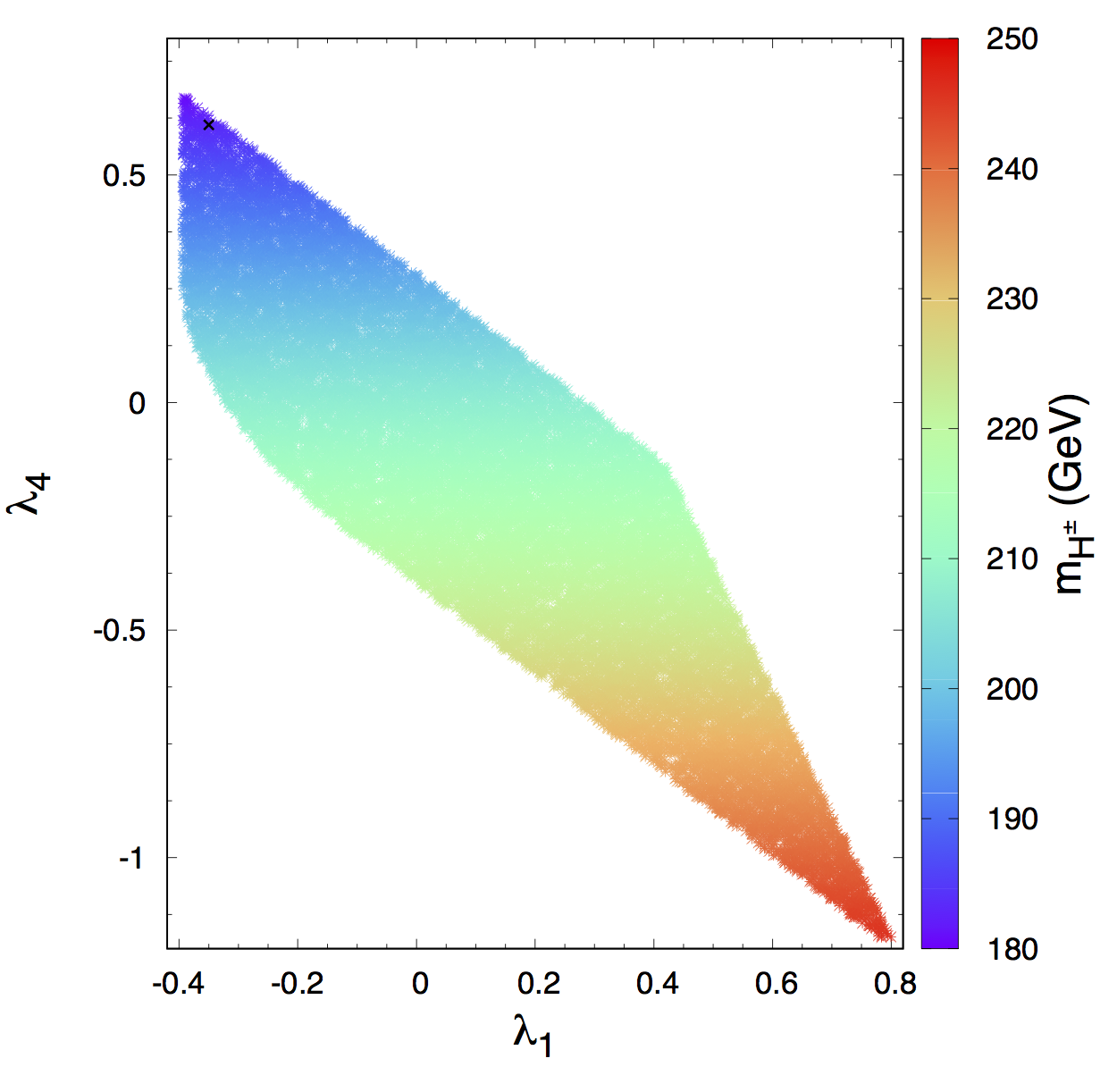}}
\end{center}
\end{minipage}
\\
\begin{minipage}{6.1cm}
\begin{center}
\resizebox{1.1\textwidth}{!}{\includegraphics{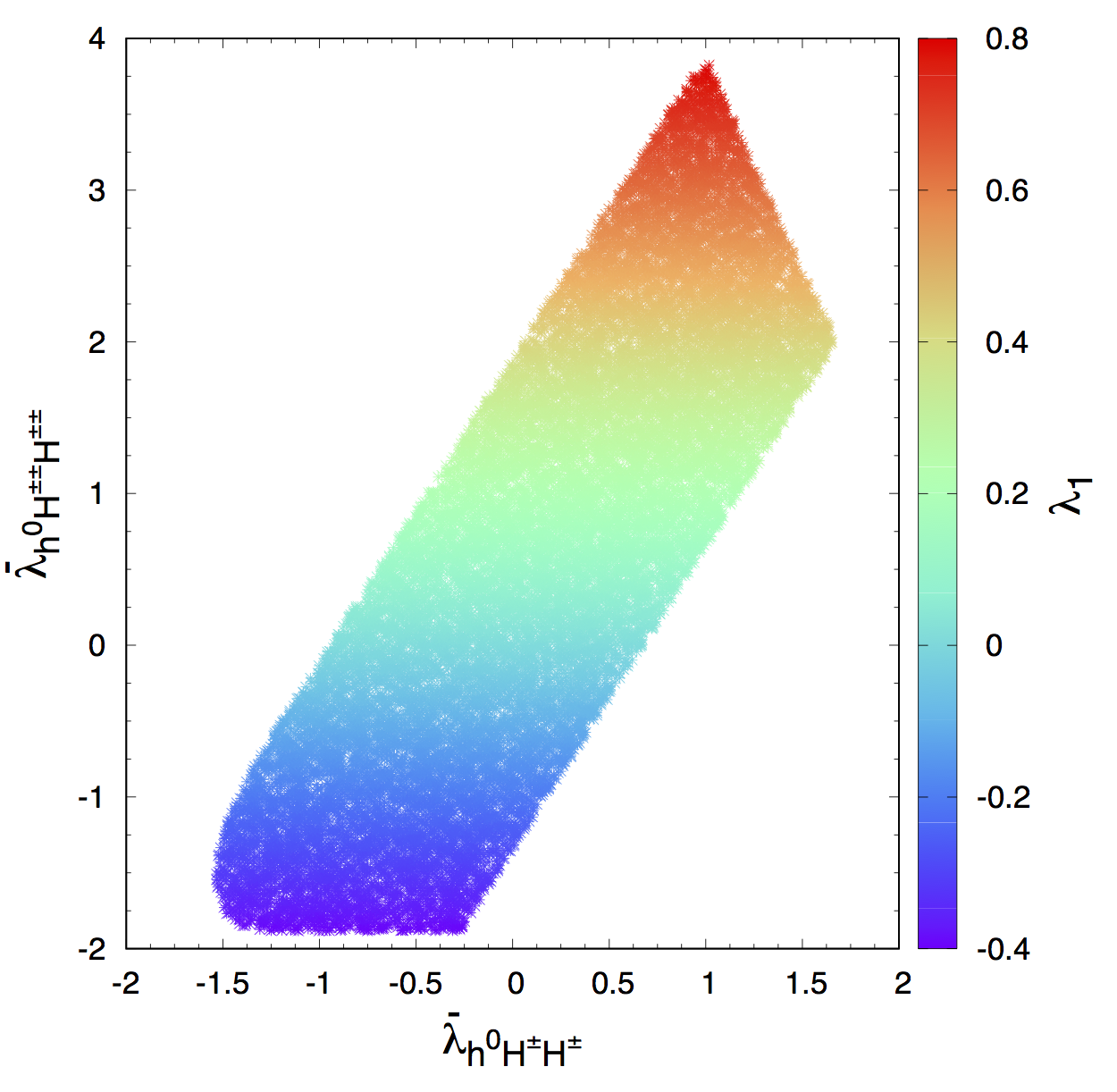}}
\end{center}
\end{minipage}
\hspace{0.6cm}
\begin{minipage}{6.1cm}
\begin{center}
\resizebox{1.1\textwidth}{!}{\includegraphics{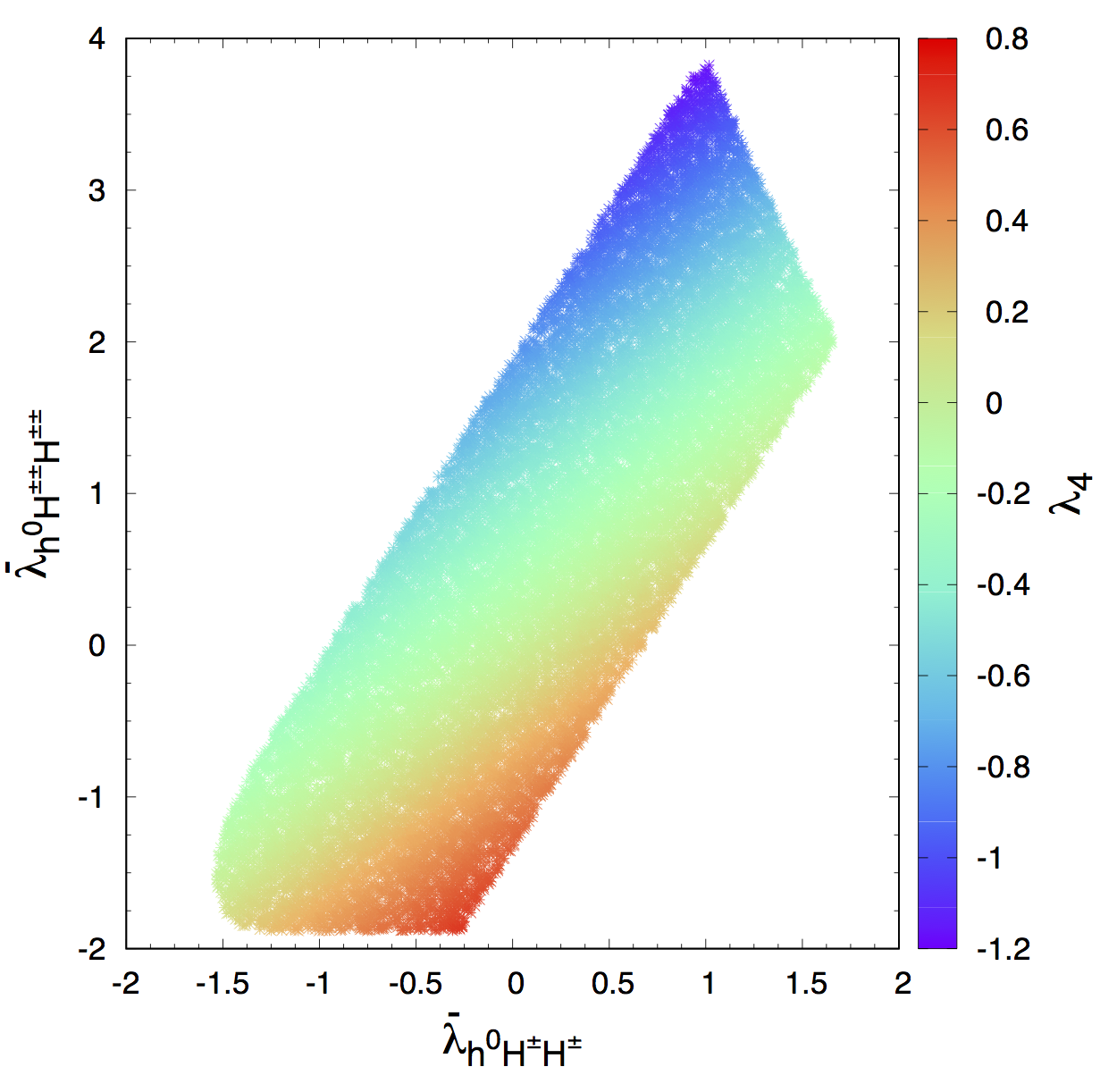}}
\end{center}
\end{minipage}
\caption{Upper panel : the allowed parameter space of the HTM given by the variation of $m_{H^{\pm\pm}}$ (left) and $m_{H^\pm}$ (right) in $(\lambda_1,\lambda_4)$ plane. In the lower panel: the correlation between $\bar{\lambda}_{h^0 H^{+}H^{-}}$ 
and $\bar{\lambda}_{h^0 H^{++}H^{--}}$ following the sign of $\lambda_1$ (left) and $\lambda_4$ (right). Input parameters are $\lambda=0.517$ ($m_{h^0}=125.09$ GeV), $\lambda_3=2\lambda_2=0.2$, $v_\Delta=\mu = 1$ GeV.}
\label{fig:fig2}
\end{figure*}

We first comment about $e^+e^-\to \gamma H$ in the SM. Like $H \to \gamma\gamma$ and $H \to \gamma Z$, the vertex contribution in $e^+e^-\to \gamma H$ is dominated by the W loops  while the top contribution is sub-leading and 
interfere destructively with the W loops \cite{Kanemura:2018esc}. 
For center of mass energy $\leq 350$ GeV, the SM box contribution is almost of the same order as 
the vertex and interfere destructively while for  higher cm energy $\geq 350$ GeV the total cross section is dominated by 
the boxes and are constructive with the vertex.

\noindent 
We start our analysis by emphasizing the impact of the new field $\Delta$. For that, we perform a scan over all the allowed parameters spac, setting $h^0$ to mimic the observed 125 GeV Higgs boson at the LHC, and taking into account above all theoretical and experimental constraints. It is worth mentioning that the virtual effects of $H^{\pm}$ and $H^{\pm\pm}$ states in $h^0\gamma\gamma$ and $h^0\gamma Z$ couplings bring in a high sensitivity to $\lambda_1$ and $\lambda_{14}^+$ respectively. This is because the dependence of 
$h^0 H^\pm H^\mp$ and $ h^0H^{\pm\pm}H^{\mp\mp}$ triple couplings on $\lambda_1$ and $\lambda_{14}^+$ comes with the large doublet {\it vev} while the one of $\lambda_2$ and $\lambda_3$ are associated with the small triplet {\it vev}, see eqs.(\ref{eq:redgcalHHpp}) and (\ref{eq:redgcalHHp}). 
Therefore, the sensitivity to $\lambda_{2,3}$ in the processes under study is marginal.

In Fig.\ref{fig:fig2} above, we exhibit the allowed range for $(\lambda_1,\lambda_4)$ as well as the size of the triple couplings $\bar{\lambda}_{h^0 H^{+}H^{-}}$, $\bar{\lambda}_{h^0 H^{++}H^{--}}$.
Indeed, the upper panel displays a strong dependency between $\lambda_{1,4}$ couplings and charged Higgs boson masses; $m_{H^\pm}$ (left) and $m_{H^{\pm\pm}}$ (right).  It is clear from this plot the range allowed for $-0.4<\lambda_1<0.8$ is rather limited. This is mainly due to the
BFB constraints combined with light spectrun $m_{H^\pm}, m_{H^{\pm\pm}}, m_A <280$ GeV.
It is clear that for$\lambda_1>0$ and $\lambda_4<0$, $m_{H^\pm}$ (resp $m_{H^{\pm\pm}}$) takes its larger values which stands below $250$ GeV (resp 280 GeV). 

The lower panel shows that, according to Eq.(\ref{eq:redgcalHHpp}), the coupling $\bar{\lambda}_{h^0 H^{\pm\pm}H^{\mp\mp}}$ 
is proportional to $\lambda_1$. Such a fact is visible in the left-side where one can see that 
the sign of $\bar{\lambda}_{h^0 H^{\pm\pm}H^{\mp\mp}}$  is completely dictated by $\lambda_1$ sign.
The situation is quite different for $\bar{\lambda}_{h^0 H^{\pm}H^{\mp}}$ which could have both signs for positive or negative
$\lambda_1$, see Eq.(\ref{eq:redgcalHHp}). However,  $\bar{\lambda}_{h^0 H^{\pm}H^{\mp}}$ coupling have different texture as a function of $\lambda_1$ and $\lambda_4$. 
As it can be seen, for large and negative $\lambda_4$, the coupling $\bar{\lambda}_{h^0 H^{\pm}H^{\mp}}$
 is maximal and positive and flip sign for a wide range of large and positive $\lambda_4$.
Thus, it can be said that $\lambda_1$ and $\lambda_4$ are important parameters of the model, and more severely restrictions on their values from analysis involving naturalness have been studied in Ref.\cite{chabab2016}. 
\begin{figure*}[!ht]
\begin{minipage}{5.4cm}
\begin{center}
\resizebox{1.12\textwidth}{!}{\includegraphics{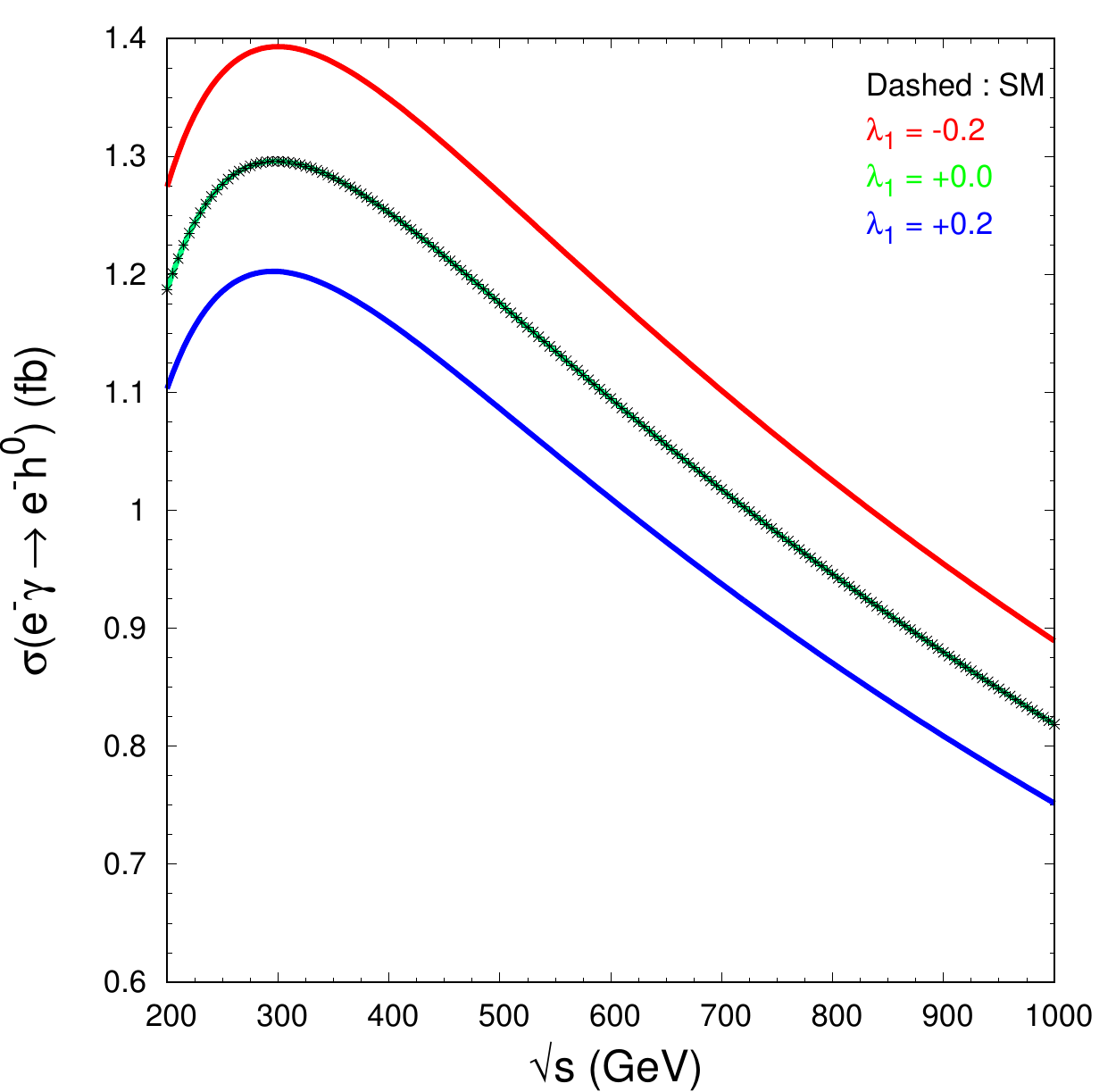}}
\end{center}
\end{minipage}
\hspace{0.4cm}
\begin{minipage}{5.4cm}
\begin{center}
\resizebox{1.12\textwidth}{!}{\includegraphics{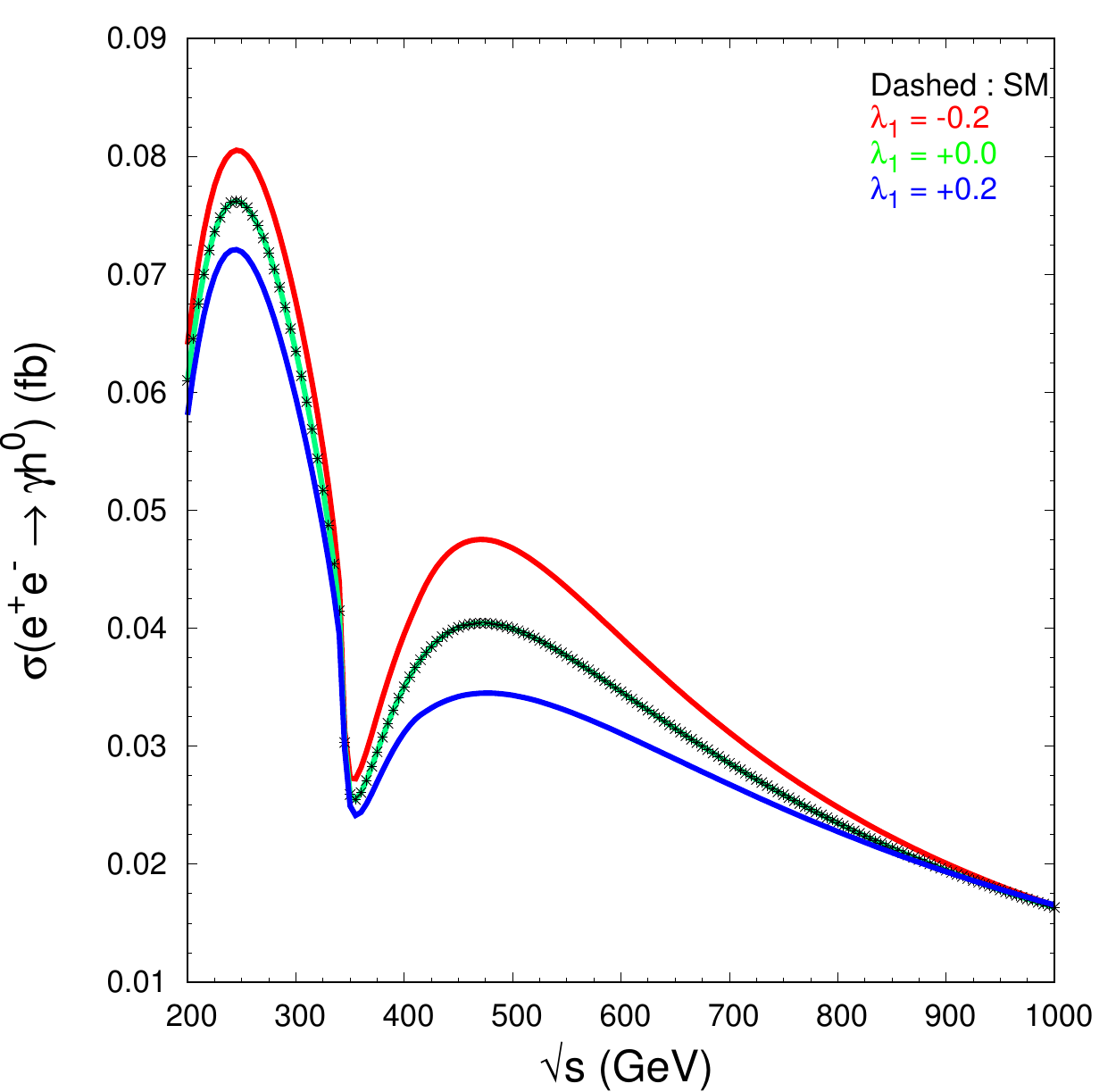}}
\end{center}
\end{minipage}
\hspace{0.4cm}
\begin{minipage}{5.4cm}
\begin{center}
\resizebox{1.08\textwidth}{!}{\includegraphics{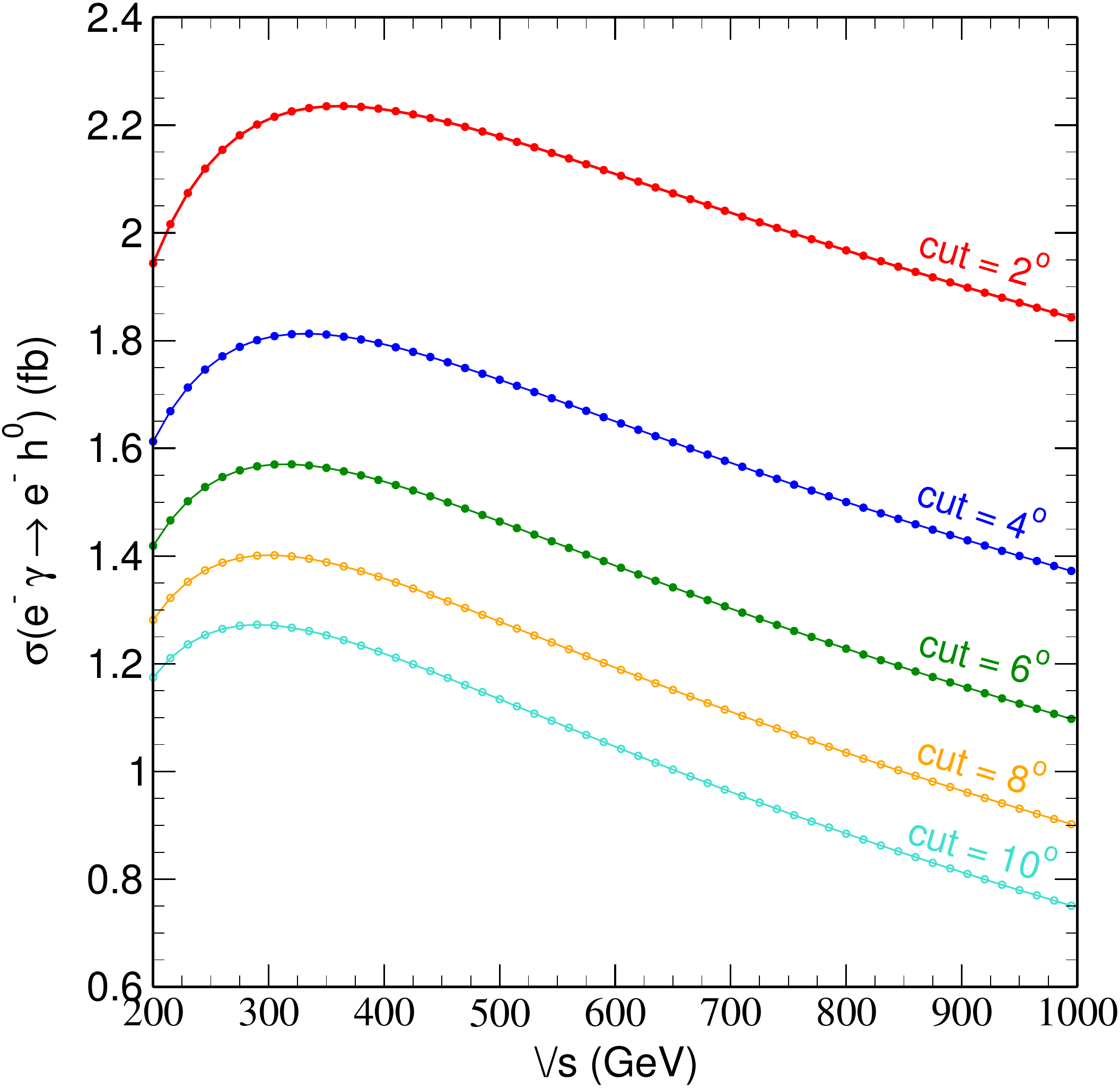}}
\end{center}
\end{minipage}
\caption{Total unpolarized cross section in {\it fb} for $e^+e^- \to \gamma h^0$ (left) and $e^-\gamma \to e^- h^0$ (middle) as a function of center-of-mass energy for various values of $\lambda_1$ and $\lambda=0.517$ ($m_{h^0}=125.09$ GeV), $\lambda_3=2\lambda_2=0.2$, $v_\Delta=\mu = 1$ GeV and $\lambda_4=0$. In the right side the total cross section $\sigma(e^-\gamma \to e^- h^0)$ in {\it fb} is displayed as a function of $\sqrt{s}$ for five different angle cuts of $2^{o}$, $4^{o}$, $6^{o}$, $8^{o}$ and $10^{o}$ with $\lambda_1=-0.2$.}
\label{fig:fig1}
\end{figure*}

In Fig.\ref{fig:fig1} we would like to establish a benchmark for comparing both unpolarized cross section $e^+e^-\to h^0\gamma$ and $e^- \gamma \to e^- h^0$  as a function of center-of-mass energy $\sqrt{s}$ with respect to SM one. 
For this purpose, we also draw the SM values for the cross sections as a function of cm energy. We stress here, as we will see later, 
the amplitude of the t-channel contribution to $e^-\gamma \to h^0 e^-$ contains a singularity when $\cos\theta \approx 1$, i.e when the angle between the incoming and outgoing electrons vanish. Therefore, we introduce a cut of 10$^\circ$ 
when we evaluate the total cross section for $e^-\gamma \to h^0 e^-$. 
However, in the case of $e^+e^- \to \gamma h^0$ process, we have 
checked that the total cross section does not depend on small cut. 
This is because of a cancelation between t-channel vertices and boxes diagrams.

\begin{table}[!h]
\caption{Benchmark points for Fig.\ref{fig:fig1}.}
\centering
\begin{tabular}{c|c|c|c|c}
\hline
 Bench. & $\lambda_1$ &  $\lambda_4$ & $m_{H^\pm}/{\rm GeV}$ &  $m_{H^{\pm\pm}}/{\rm GeV}$     \\ \hline 
\hline
BP1. & $-0.2$ & $0.0$ & \multicolumn{2}{|c}{\multirow{3}{*}{$\approx 206$}}  \\
BP2. & $0.0$ & $0.0$ & \multicolumn{2}{|c}{} \\
BP3. & $+0.2$ & $0.0$ & \multicolumn{2}{|c}{} \\
\hline 
\hline
\end{tabular}
\label{tab:bp_nonalignment}
\end{table}
        
As it can be noticed, the cross sections are enhanced near the region 
$\sqrt{s}\approx\,250$ GeV. As far as $\sqrt{s}$ increases further, the destructive interference between the
SM and HTM contributions gets more severe and becomes maximal near 
 the $t\bar{t}$ threshold, responsible for the dips clearly seen in the figure.
After crossing the $t\bar{t}$ threshold, a constructive or destructive interference with SM contribution 
depending on the sign of $\lambda_1$ which is the same as the sign of $\bar{\lambda}_{h^0 H^{\pm\pm}H^{\mp\mp}}$.
Note that the vertex contribution for  $e^+e^- \to h^0\gamma$ scales like 1/s 
and thus drop steeply for large $\sqrt{s}$, while for $e^-\gamma\to e^- h^0$ which have a t-channel contribution, 
the drop of the cross section for large $\sqrt{s}$ is slower than for $e^+e^-\to \gamma h^0$.

\begin{figure*}[!h]
\centering
\begin{minipage}{6.1cm}
\begin{center}
\resizebox{1.1\textwidth}{!}{\includegraphics{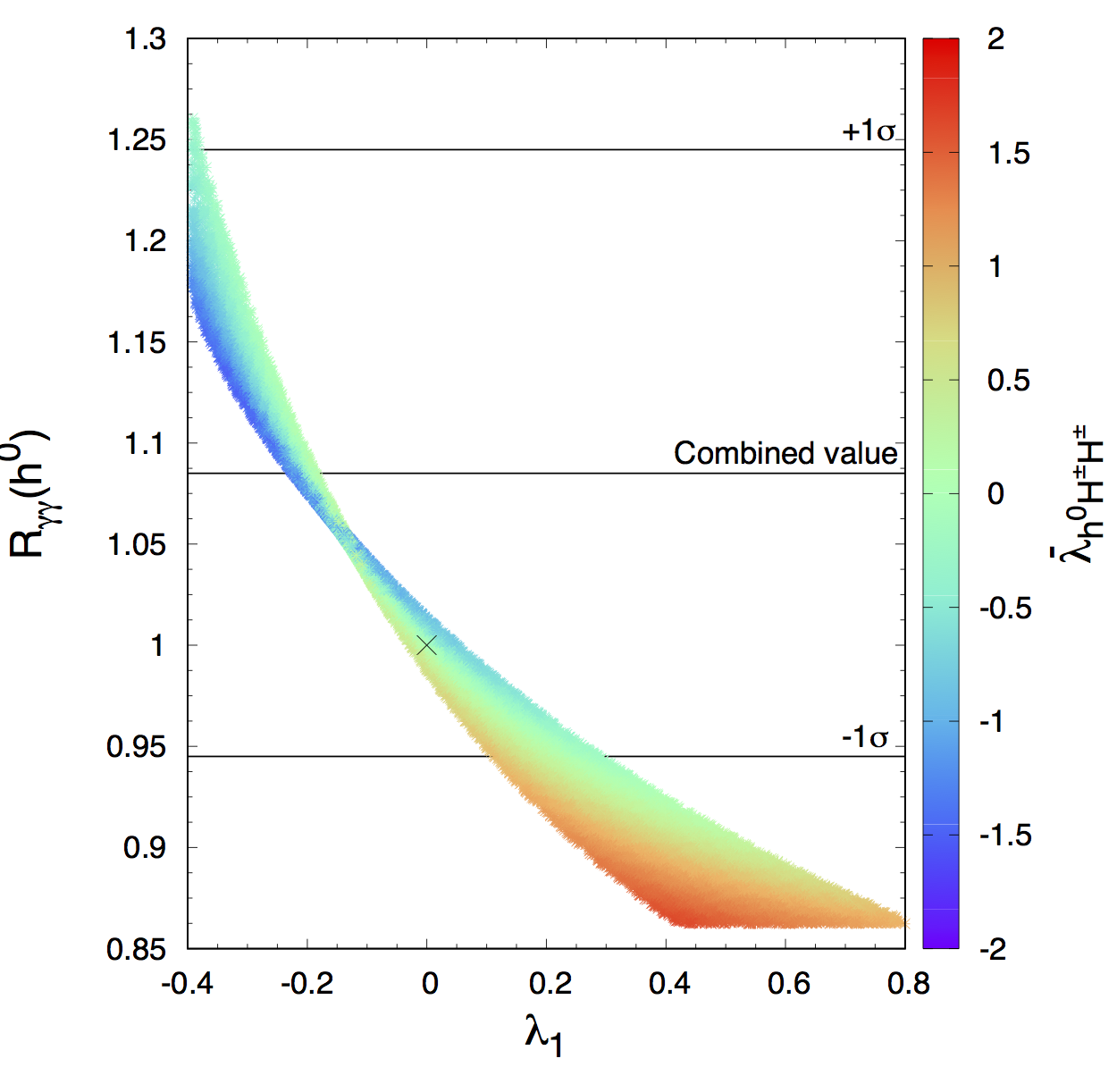}}
\end{center}
\end{minipage}
\hspace{0.6cm}
\begin{minipage}{6.1cm}
\begin{center}
\resizebox{1.1\textwidth}{!}{\includegraphics{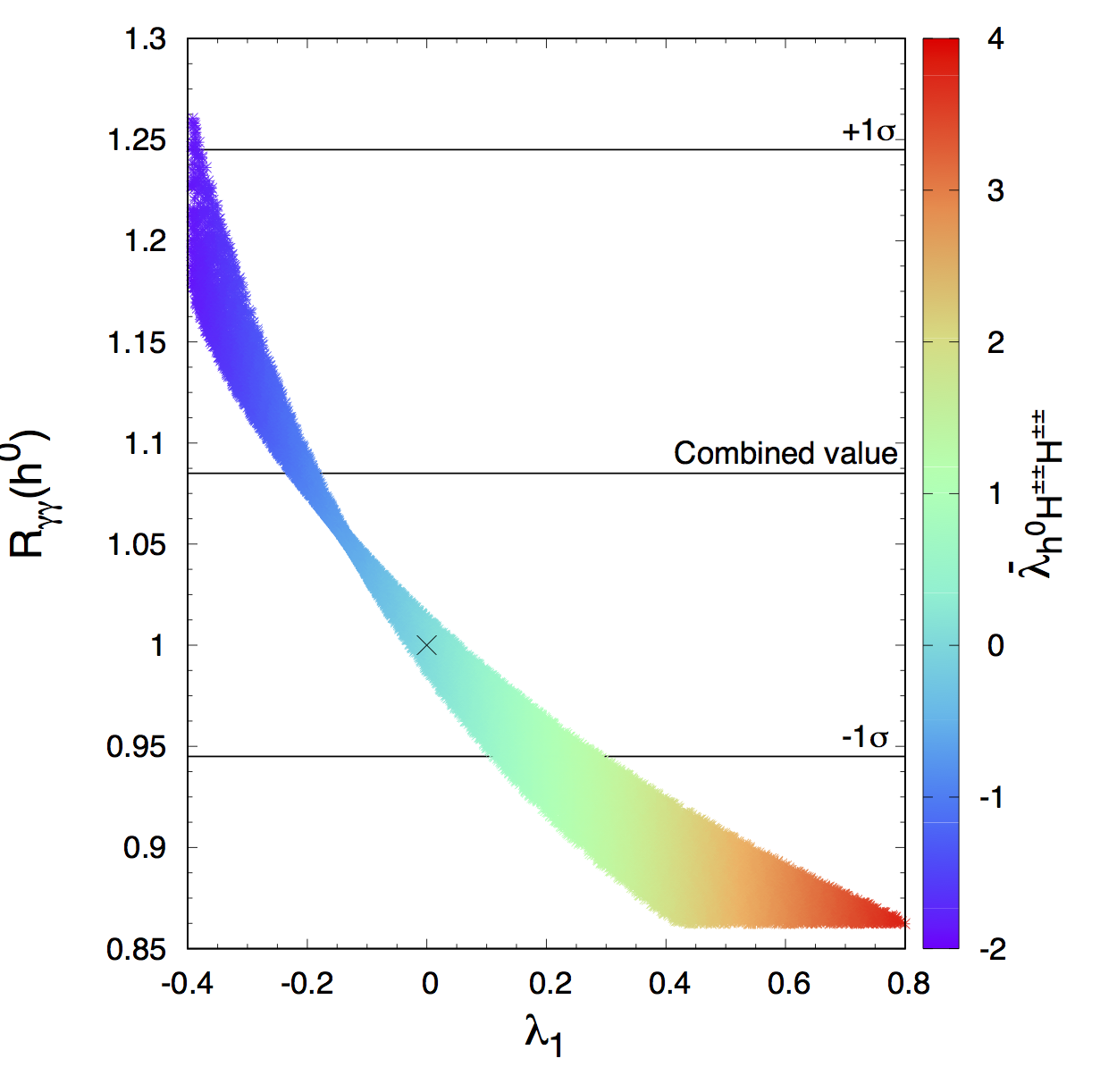}}
\end{center}
\end{minipage}
\\
\begin{minipage}{6.1cm}
\begin{center}
\resizebox{1.1\textwidth}{!}{\includegraphics{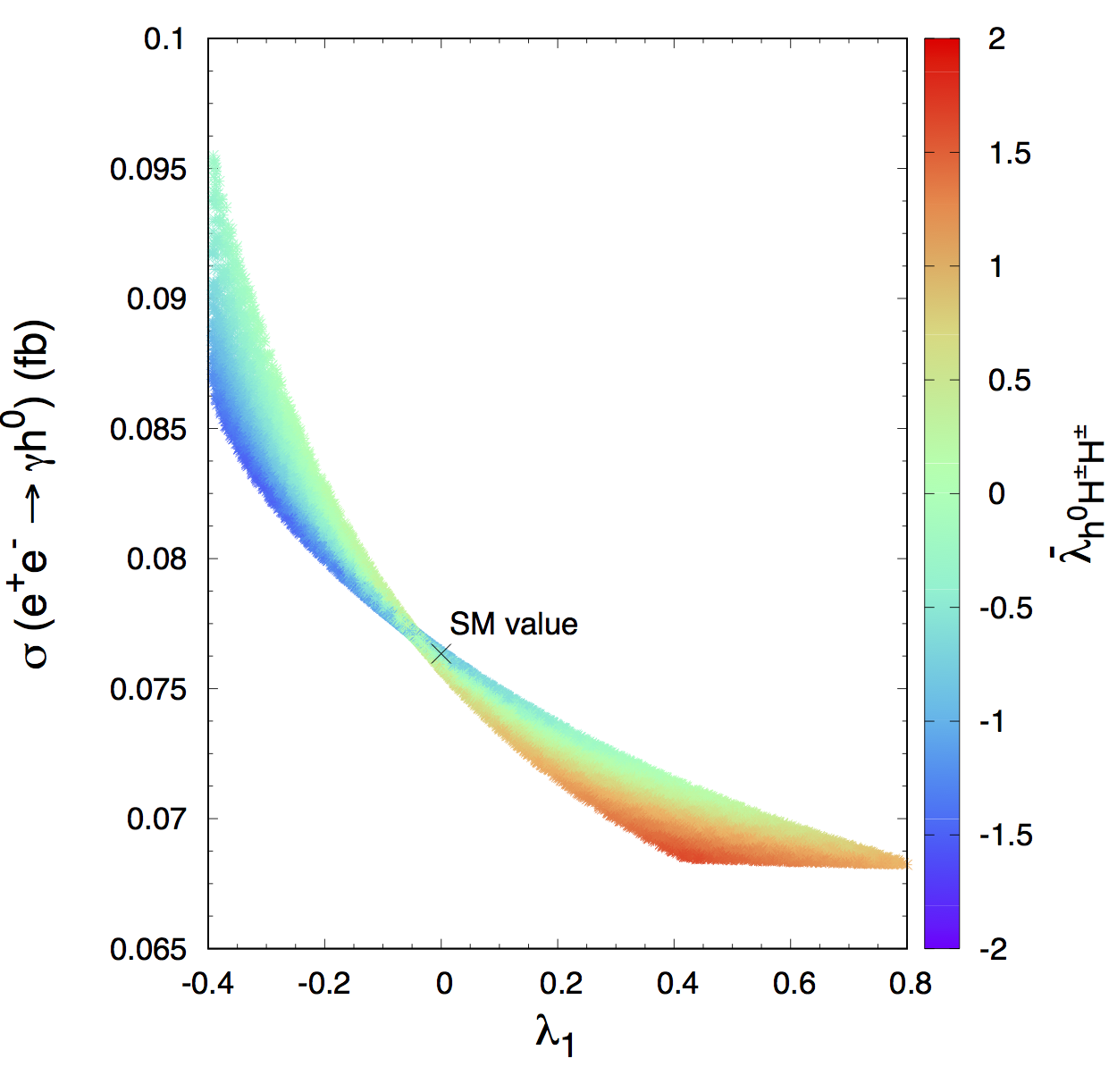}}
\end{center}
\end{minipage}
\hspace{0.6cm}
\begin{minipage}{6.1cm}
\begin{center}
\resizebox{1.1\textwidth}{!}{\includegraphics{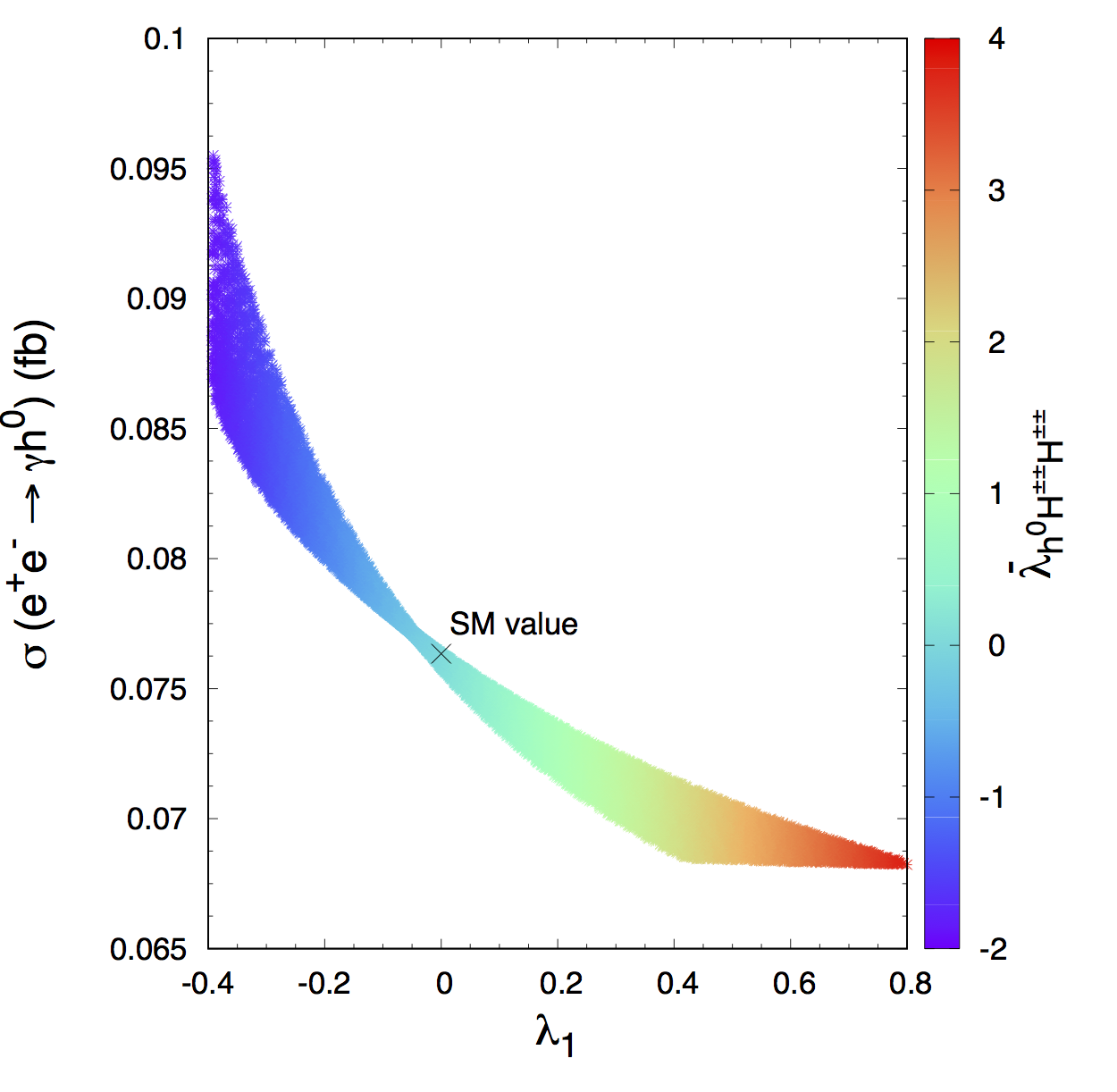}}
\end{center}
\end{minipage}
\caption{Signal strength $R_{\gamma\gamma}(h^0)$ (upper panels) and $\sigma(e^+e^- \to \gamma h^0)$ (fb) total cross section 
for $\sqrt{s}=250$ GeV (lower panel)  as a function of $\lambda_1$ coupling. The other inputs are 
$\lambda=0.517$ ($m_{h^0}=125.09$ GeV), $\lambda_3=2\lambda_2=0.2$, $-1.2 \le \lambda_4 \le 0.8$, $v_\Delta=\mu = 1$ GeV. 
Horizontal lines denote the central and $\pm1\sigma$ combined diphoton strength signal reported by ATLAS \cite{Aaboud:2018xdt} and CMS \cite{Sirunyan:2018ouh} at 13 TeV. Left
color coding shows the variation of the reduced 
trilinear coupling $\bar{\lambda}_{h^0 H^{\pm}H^{\mp}}$ while the right one shows the 
$\bar{\lambda}_{h^0 H^{\pm\pm}H^{\mp\mp}}$.}
\label{fig:fig3}
\end{figure*}

Concerning the effect of charged Higgses on $R_{\gamma\gamma}(h^0)$ and the cross section $\sigma(e^+e^- \to \gamma\,h^0)$ and $\sigma(e^-\gamma \to e^-\,h^0)$, we illustrate in Fig.\ref{fig:fig3}  the variation of those observable as a function of the parameter $\lambda_1$ and the reduced couplings $\bar{\lambda}_{h^0 H^{\pm}H^{\mp}}$ and $\bar{\lambda}_{h^0 H^{\pm\pm}H^{\mp\mp}}$. Hence, the sensitivity to $\lambda_1<0$ as well as 
$\bar{\lambda}_{h^0 H^{\pm}H^{\mp}}<0$ and $\bar{\lambda}_{h^0 H^{\pm\pm}H^{\mp\mp}}<0$
is particularly striking, as one can see for  $m_{H^\pm}$ above 150 GeV, $R_{\gamma\gamma}$ is enhanced substantially beyond its SM values. This is due to a  constructive interference between the SM contribution dominantly by $W^+$, 
and that of singly and doubly charged Higgs bosons in $h^0\gamma\gamma$ and $h^0\gamma Z$ couplings. \\
On the other hand, note that both vertices given in Eqs.(\ref{eq:redgcalHHp}) and (\ref{eq:redgcalHHpp}) contribute to
 both $R_{\gamma\gamma}$ and $\sigma(e^+e^- \to \gamma\,h^0)$.  Therefore, we expect that 
 $\sigma(e^+e^- \to \gamma\,h^0)$ would vary in a similar way as $R_{\gamma\gamma}$. 
This fact is clearly seen from the lower panels of Fig.\ref{fig:fig3}.
Indeed, the cross section can reach values as high as $0.096$ fb for negative values of $\lambda_1$, 
requiring a rather light singly (and/or doubly) charged Higgs bosons in the range $[100, 200]$ GeV. 
Such large enhancement is related to the sensitivity to trilinear reduced couplings $\bar{\lambda}_{h^0 H^{\pm}H^{\mp}}$ and $\bar{\lambda}_{h^0 H^{\pm\pm}H^{\mp\mp}}$, which also depend on $\lambda_{1,4}$ signs.

\begin{figure*}[!h]
\centering
\begin{minipage}{6.1cm}
\begin{center}
\resizebox{1.1\textwidth}{!}{\includegraphics{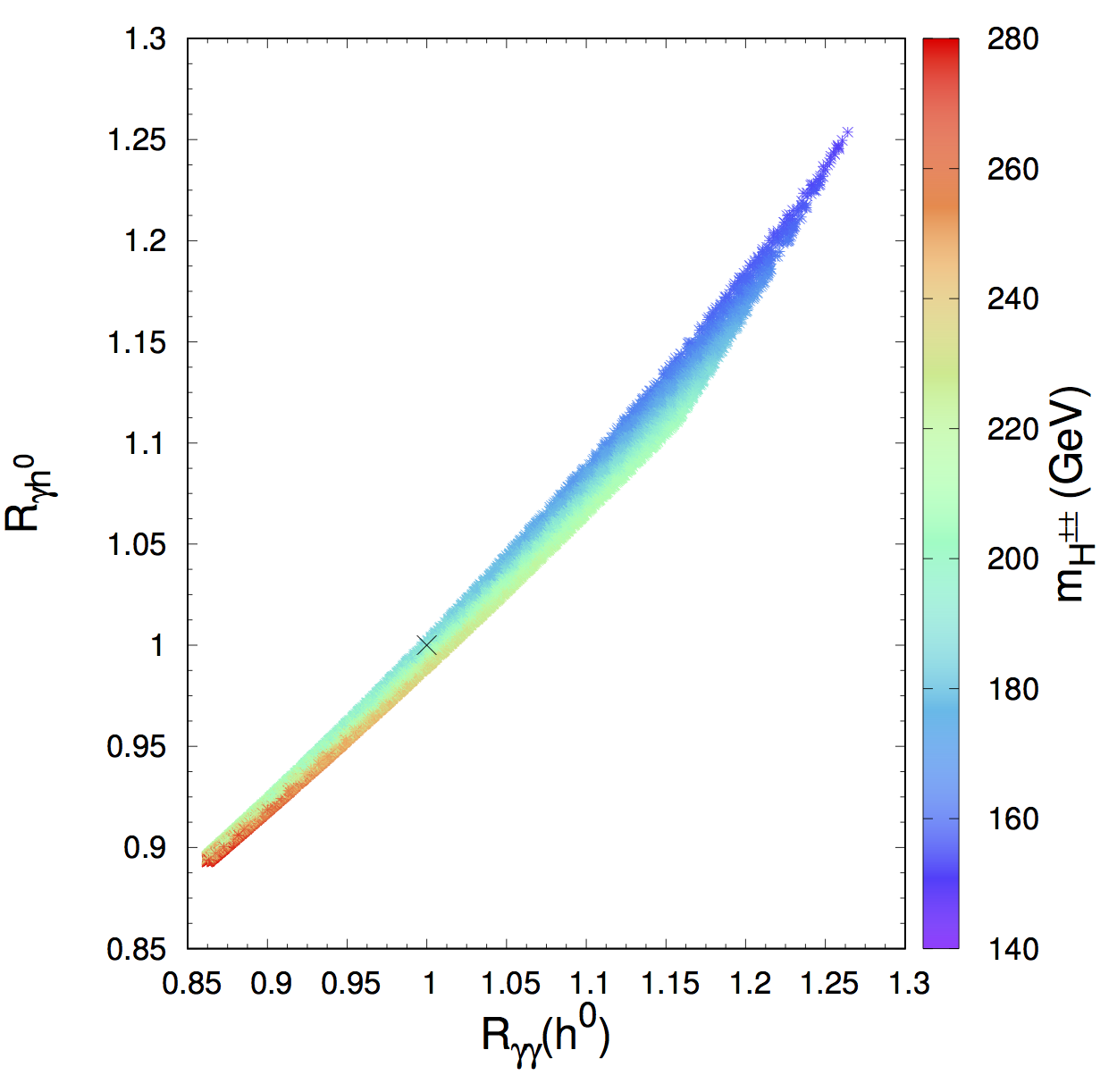}}
\end{center}
\end{minipage}
\hspace{0.6cm}
\begin{minipage}{6.1cm}
\begin{center}
\resizebox{1.1\textwidth}{!}{\includegraphics{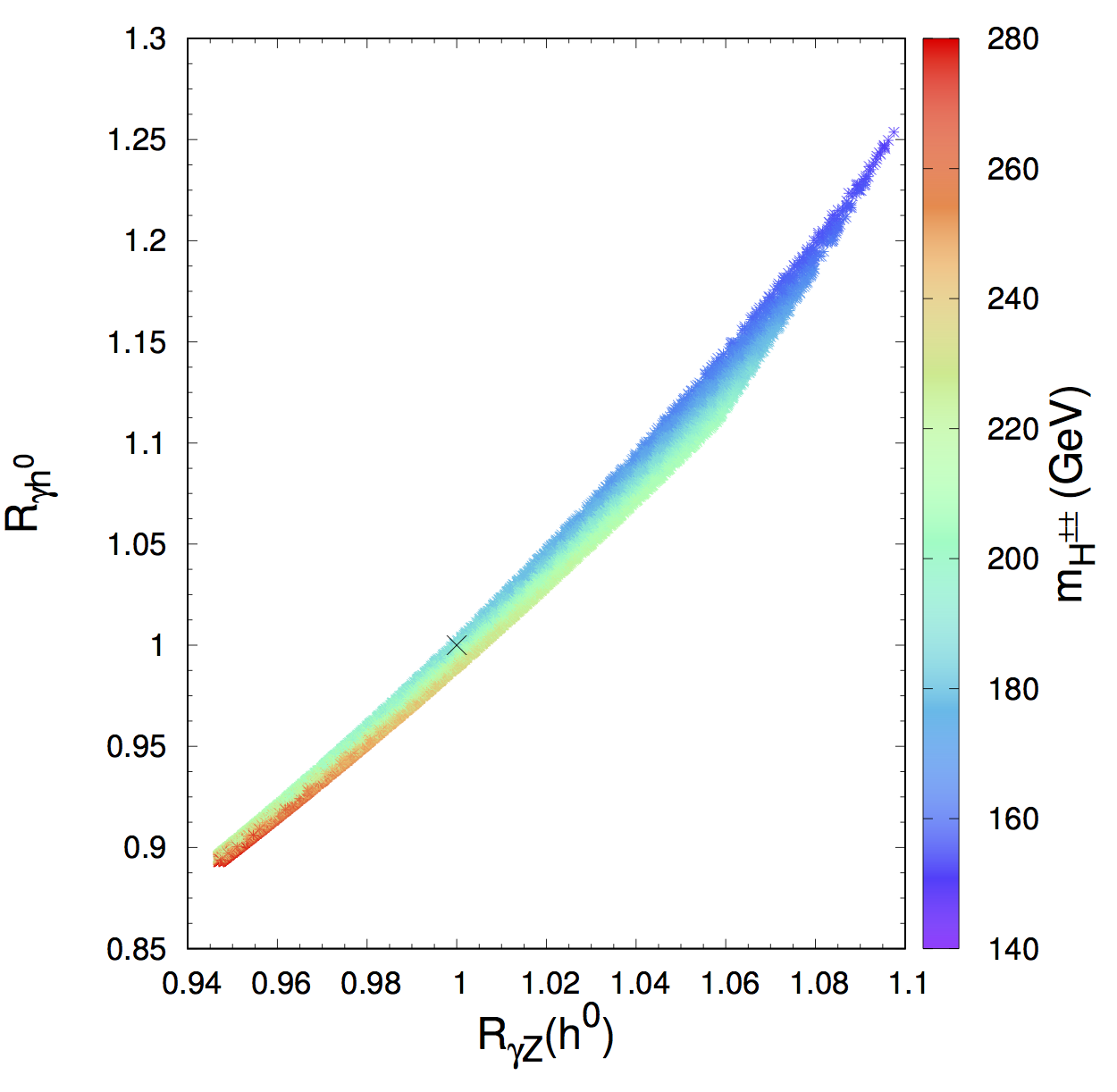}}
\end{center}
\end{minipage}
\caption{$R_{\gamma h^0}$ correlation with $R_{\gamma\gamma}(h^0)$ (left) and $R_{\gamma\,Z}(h^0)$ (right) in the HTM. Color coding shows the variation of $m_{H^{\pm\pm}}$. Inputs as in Fig.\ref{fig:fig3}
}
\label{fig:fig4}
\end{figure*}

\begin{figure*}[!h]
\centering
\begin{minipage}{6.1cm}
\begin{center}
\resizebox{1.1\textwidth}{!}{\includegraphics{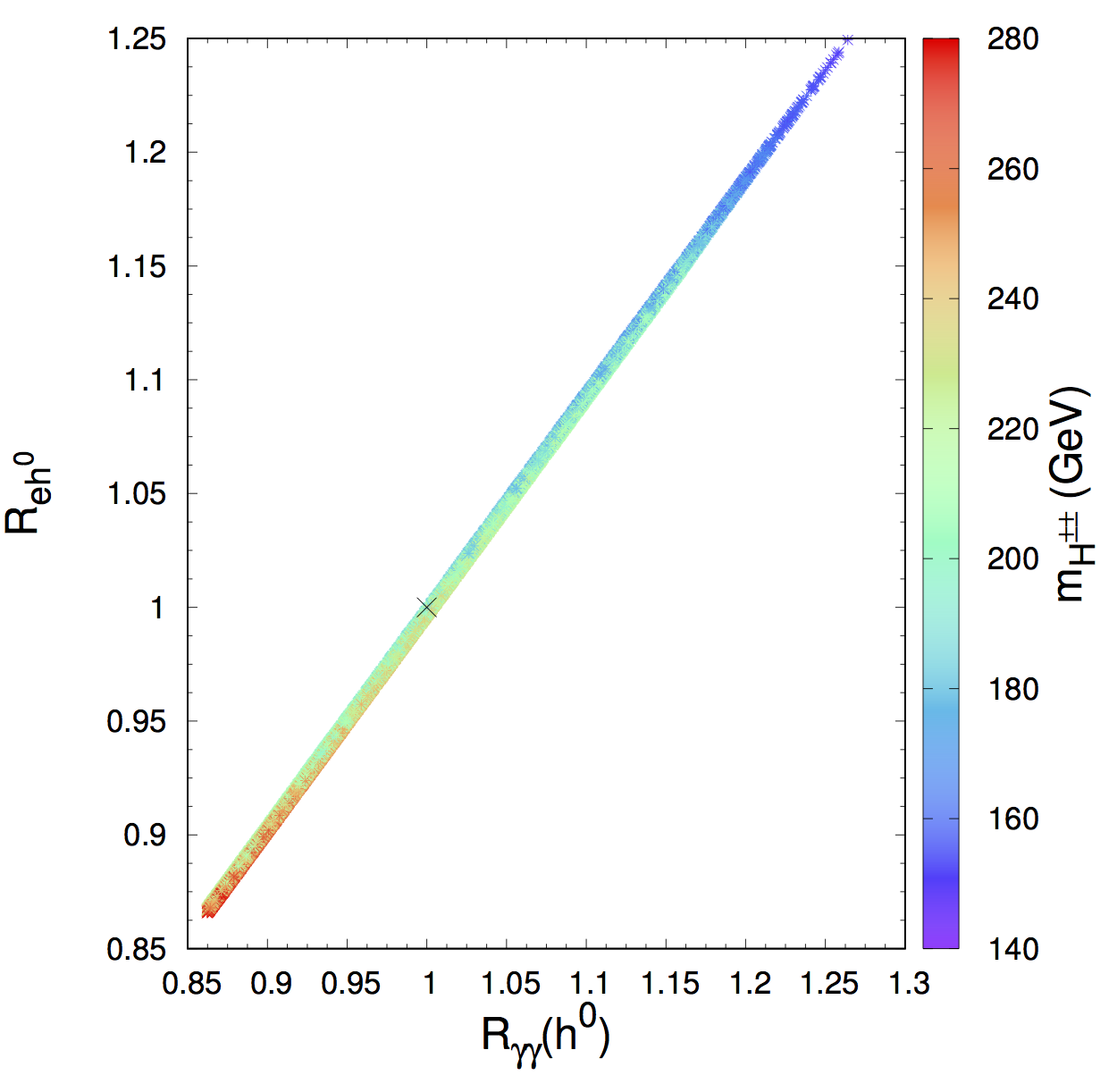}}
\end{center}
\end{minipage}
\hspace{0.6cm}
\begin{minipage}{6.1cm}
\begin{center}
\resizebox{1.1\textwidth}{!}{\includegraphics{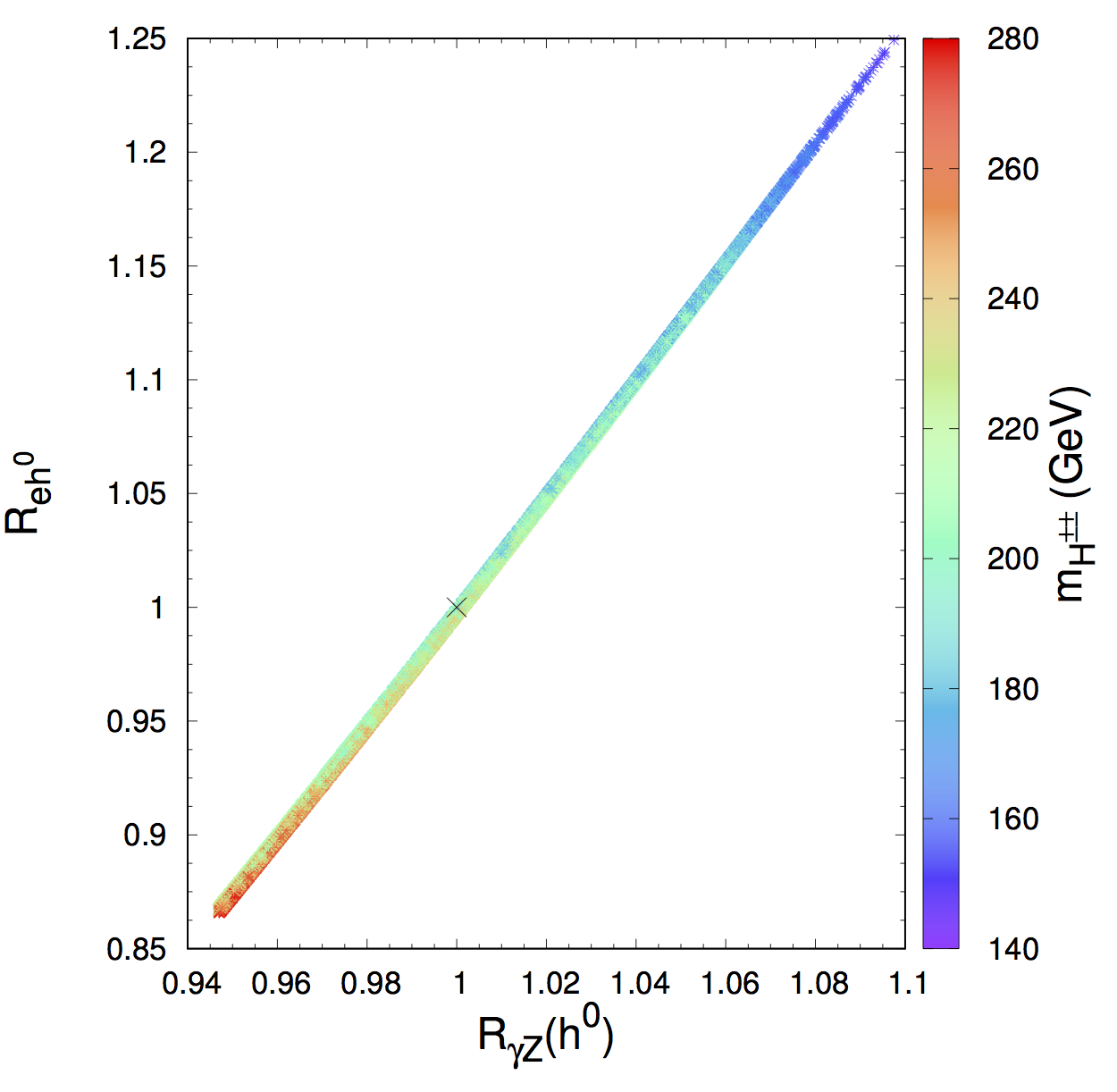}}
\end{center}
\end{minipage}
\caption{$R_{e^- h^0}$ correlation with $R_{\gamma\gamma}(h^0)$ (left) and $R_{\gamma\,Z}(h^0)$ (right) in the HTM. Color coding shows the variation of $m_{H^{\pm\pm}}$. Inputs as in Fig.\ref{fig:fig3}}
\label{fig:fig44}
\end{figure*}

It is, therefore, possible to address a possible correlation between $R_{\gamma\,V}$, $V=\gamma, Z$, and $R_{\gamma h^0}$ (resp. $R_{\gamma\,V}$ and $R_{e^-h^0}$) as displayed in Fig.\ref{fig:fig4} (resp. Fig.\ref{fig:fig44}). 
Depending on the parameter space, one can predict a positive correlation between these two observable. 
Thus, it can be seen clearly from these plots that when $R_{\gamma\,V}(h^0)>1$ we almost have $\sigma(e^+e^- \to \gamma\,h^0)>\sigma(e^+e^- \to \gamma\,H^{SM})$ (resp. $\sigma(e^-\gamma \to e^-h^0)>\sigma(e^-\gamma \to e^-H^{SM})$) and vice versa. 
Further improvement may be achieved, reflecting the charged Higgs masses dependence on these two observable, with regards to $\lambda_1$ and $2\lambda_1+\lambda_4$ signs, the charged Higgs loops interfere constructively (destructively) with the SM loops. 
Hence, the lighter the charged Higgs masses are, the larger the enhancement in both the total cross sections 
$\sigma(e^+e^- \to \gamma\,h^0)$, $\sigma(e^-\gamma \to e^-h^0)$ and the signal strength 
$R_{\gamma\,V}$ as illustrated in the right of Fig.\ref{fig:fig3}.

\begin{figure*}[!h]
\centering
\begin{minipage}{5.4cm}
\begin{center}
\resizebox{1.1\textwidth}{!}{\includegraphics{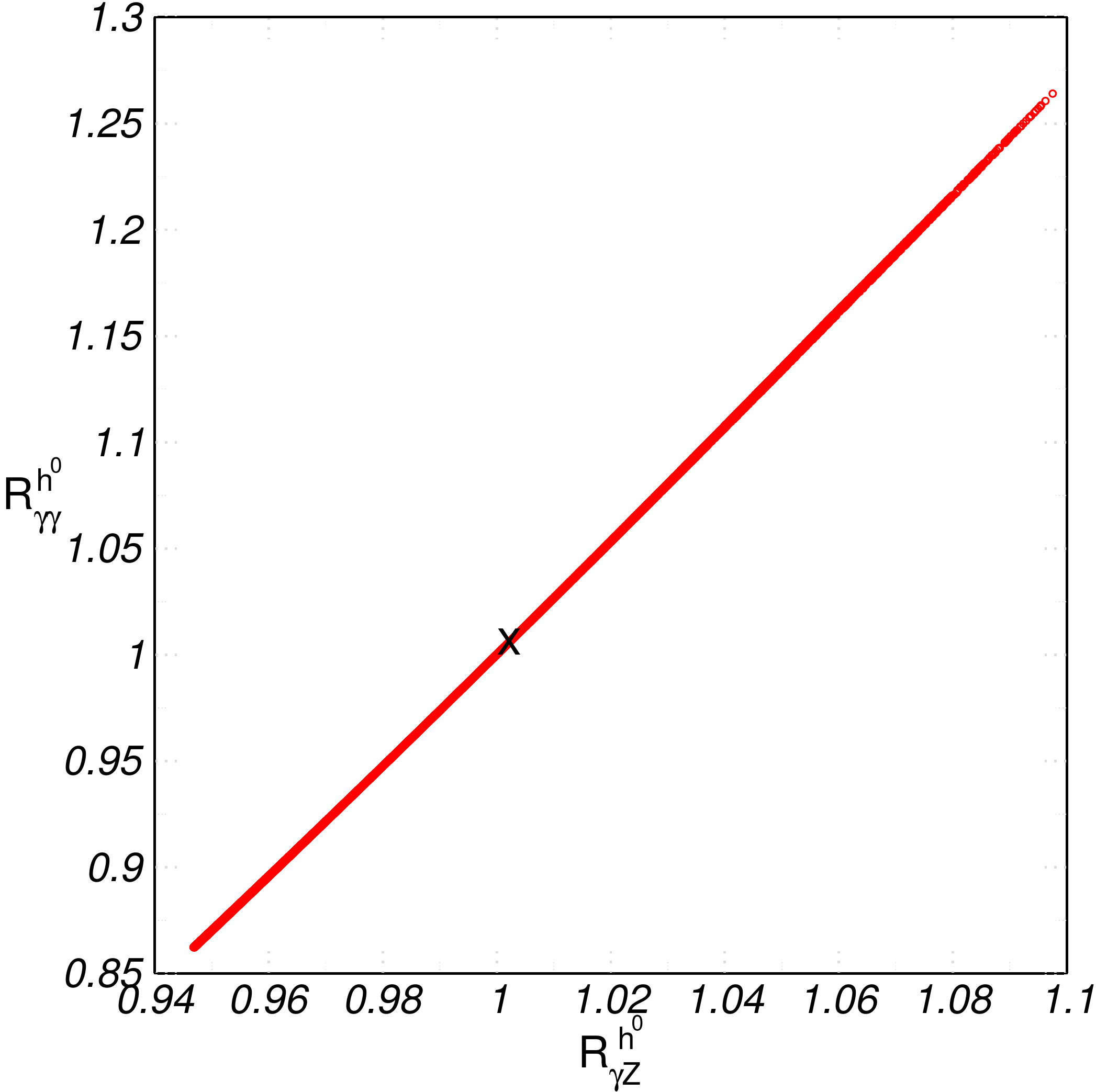}}
\end{center}
\end{minipage}
\hspace{0.4cm}
\begin{minipage}{5.4cm}
\begin{center}
\resizebox{1.095\textwidth}{!}{\includegraphics{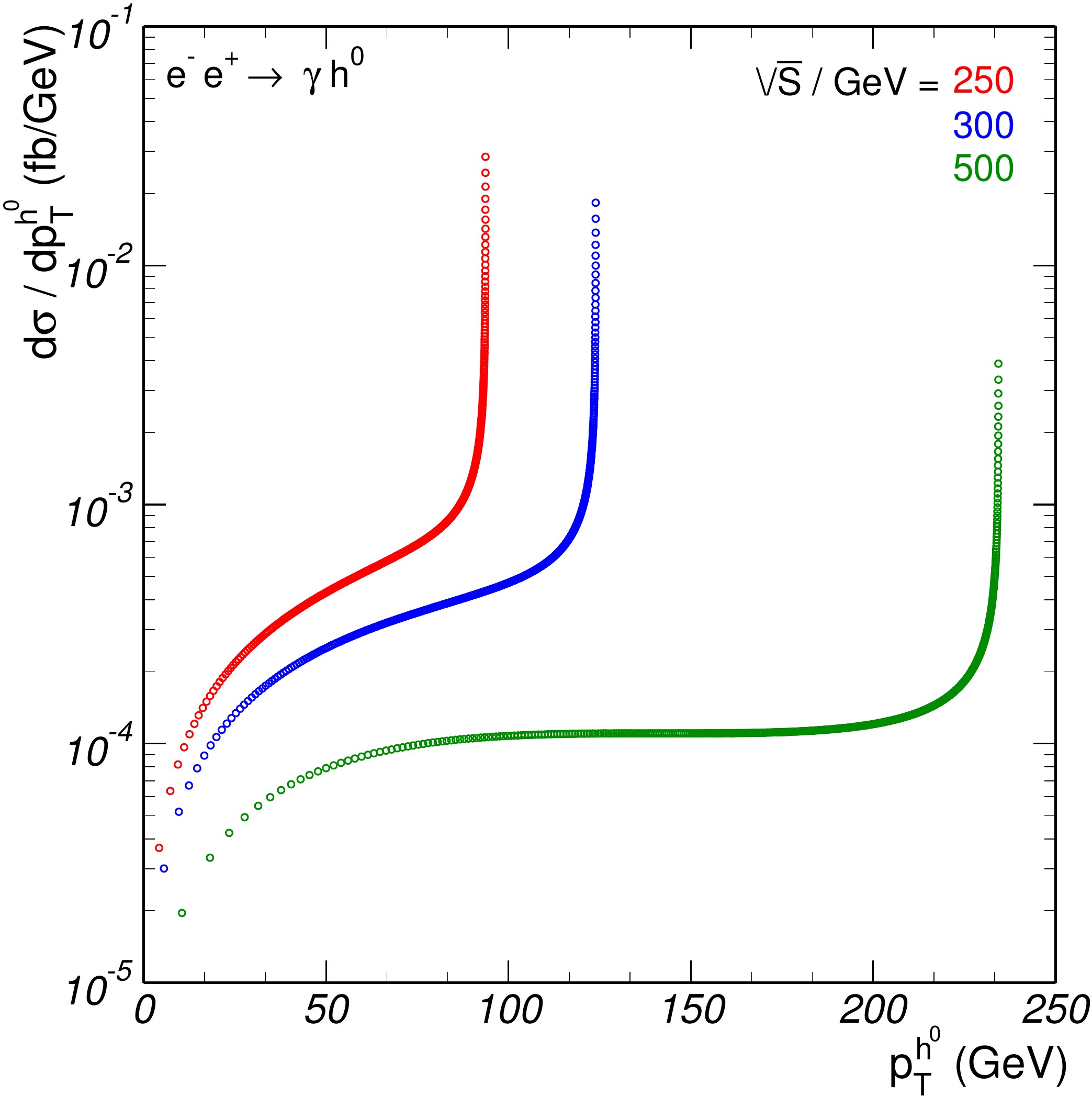}}
\end{center}
\end{minipage}
\hspace{0.4cm}
\begin{minipage}{5.4cm}
\begin{center}
\resizebox{1.095\textwidth}{!}{\includegraphics{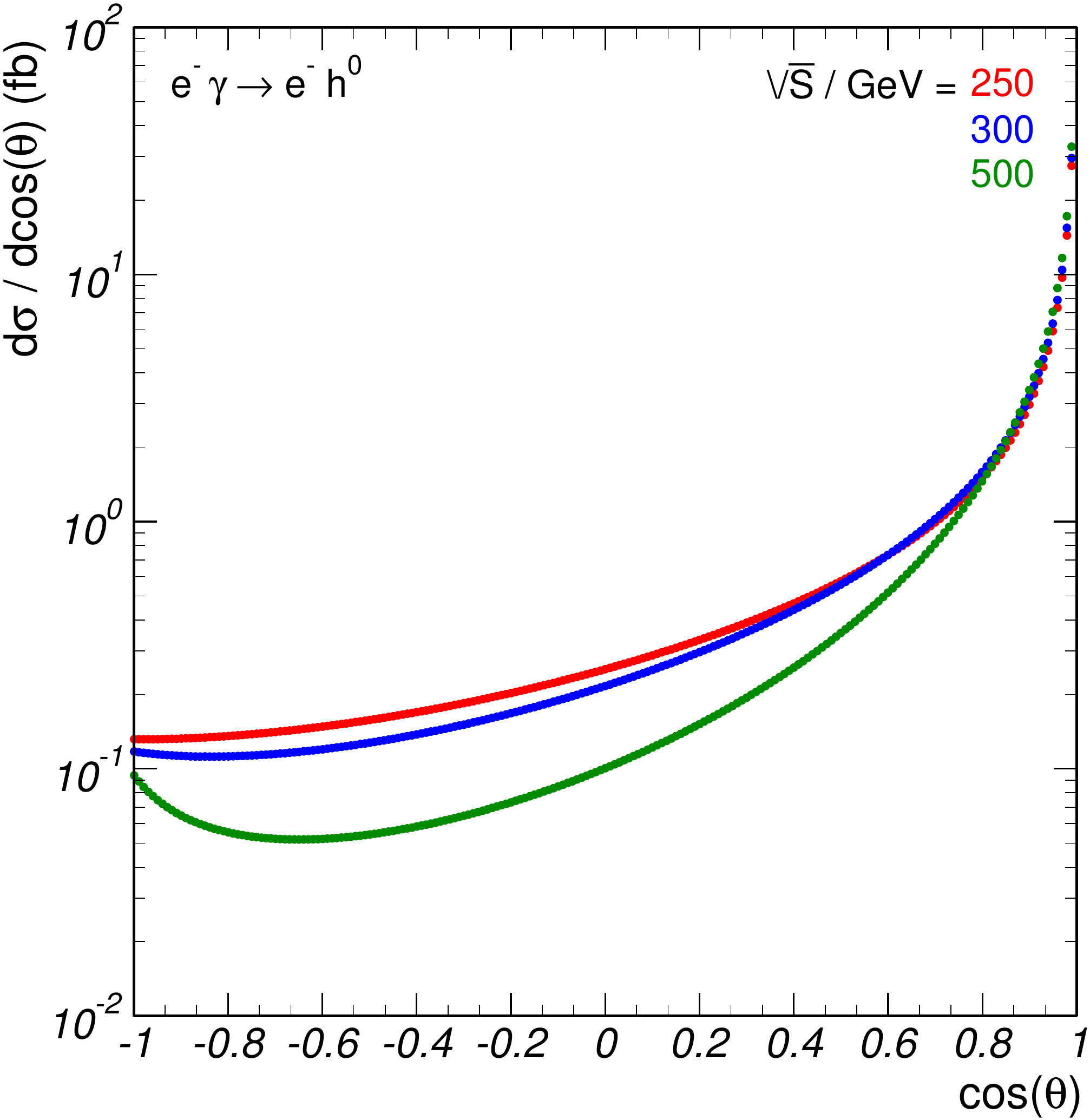}}
\end{center}
\end{minipage}
\caption{Left : $R_{\gamma\gamma}^{h^0}$ correlation with $R_{\gamma\,Z}^{h^0}$. The middle (resp. right) side shows the distribution of the transverse momentum, $p_T^{h^0}$, for the $e^-e^+ \to \gamma h^0$ (resp. differential cross section for $e^-\gamma \to e^- h^0$) for various $\sqrt{s}=$250,300 and 500 GeV, we fix here $(\lambda_1,\lambda_4)=(-0.35,+0.61)$}
\label{fig:fig5}
\end{figure*}

\begin{figure*}[!h]
\centering
\begin{minipage}{5.4cm}
\begin{center}
\resizebox{1.095\textwidth}{!}{\includegraphics{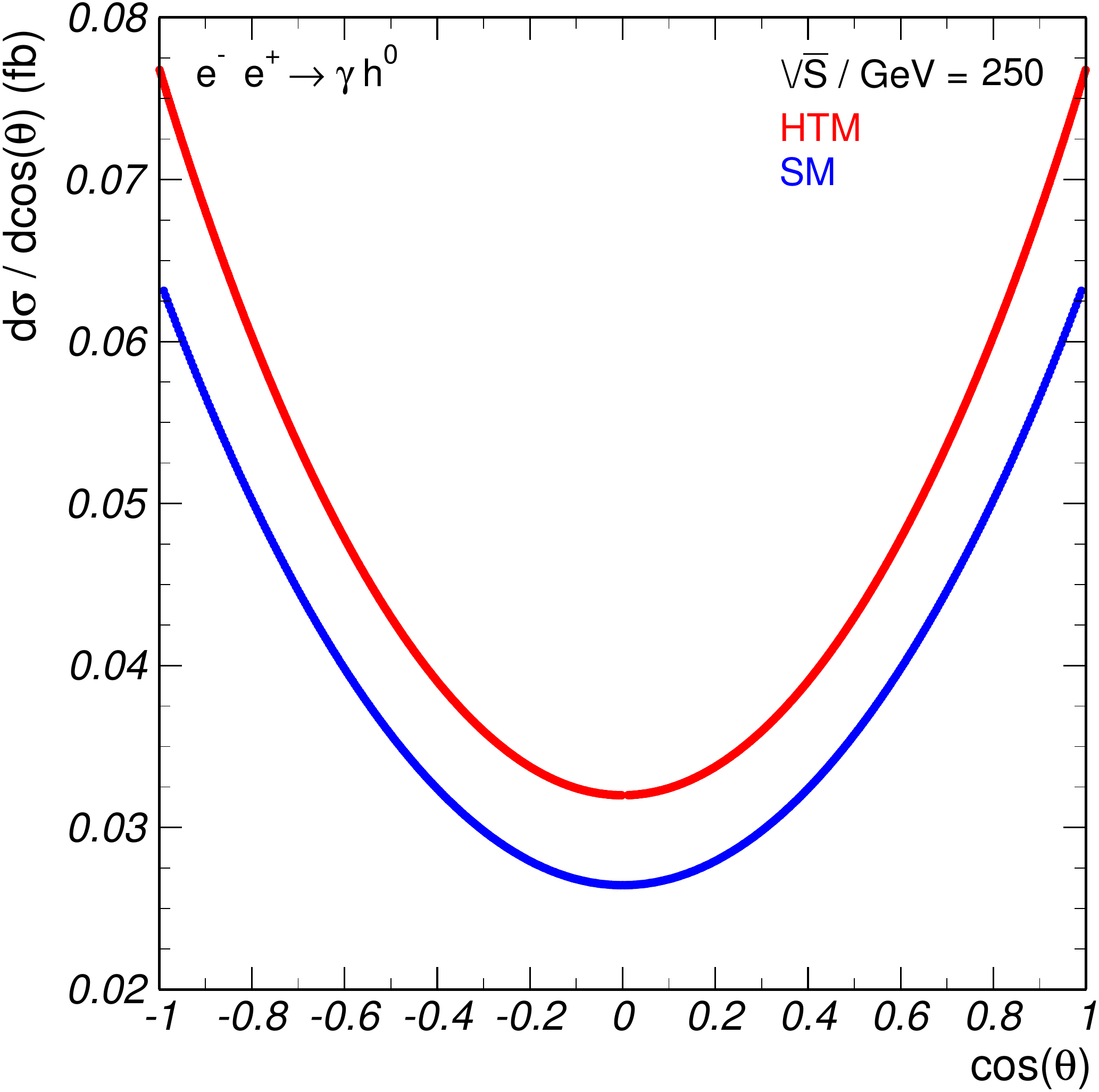}}
\end{center}
\end{minipage}
\hspace{0.4cm}
\begin{minipage}{5.4cm}
\begin{center}
\resizebox{1.095\textwidth}{!}{\includegraphics{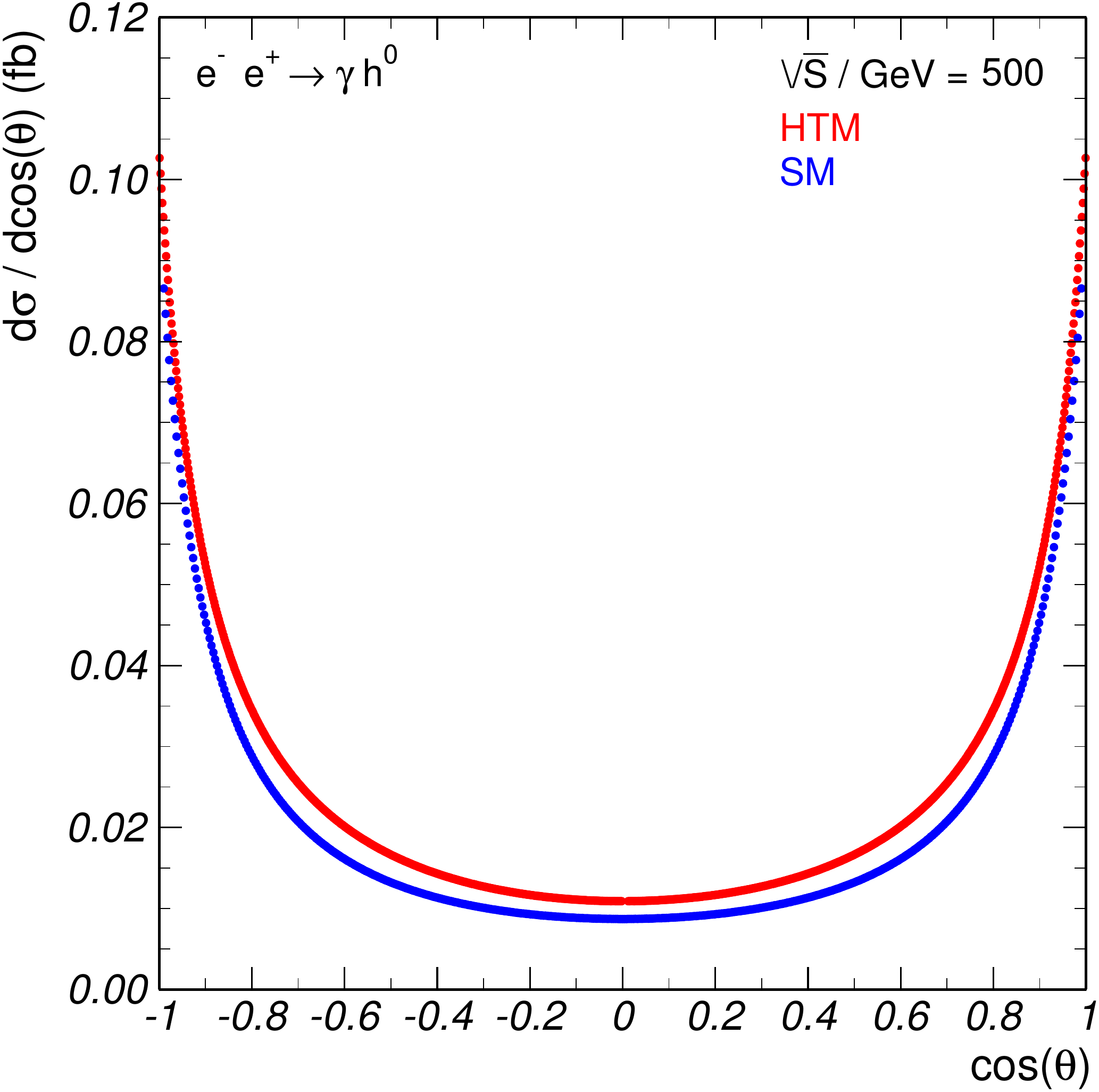}}
\end{center}
\end{minipage}
\caption{Differential cross section for $e^-e^+ \to \gamma h^0$ in both SM and HTM for two values of $\sqrt{s}=$ 250 GeV (left) and 500 GeV (right). We fix here $(\lambda_1,\lambda_4)=(-0.35,+0.61)$.}
\label{fig:fig6}
\end{figure*}

Ultimately it is interesting to understand the differential cross section for both processes $e^+e^-\to \gamma h^0$ and $e^- \gamma \to e^- h^0$, as well as the structure of the correlation between $R_{\gamma\gamma}$ and $R_{\gamma\,Z}$ for the observed $h^0$ SM-like. Such correlation is exhibited in Fig.\ref{fig:fig5} (left) for fixed values of $\mu$ and $v_\Delta$ and a scan over $\lambda_1,\lambda_4$, which demonstrate that these two decay modes are generally correlated, and typically an $R_{\gamma\gamma} \gtrsim 1$ will be consistent with the model for $R_{\gamma\,Z} \gtrsim 1$. The middle side of Fig.\ref{fig:fig5} shows the distribution for the differential cross section as a function of the transverse momentum of the Higgs boson $h^0$ ($p_T^{h^0}$), such a computation is carried out for various value of $\sqrt{s}=250, 300, 500$ GeV, employing a calculation of the one-loop scattering amplitudes for all relevant partonic  channels. It is obvious from the $p_T^{h^0}$ distribution that there is a greater likelihood of producing the $h^0$ boson with a small transverse momentum. Furthermore, the cross-section is expected to increase as the energy decrease and the increase will be faster at lower $p_T^{h^0}$ values. \\
We also show in Fig.(\ref{fig:fig5}) the differential cross section $d\sigma(e^- \gamma \to e^- h^0)/d\cos\theta$ for three center-of-mass energies $\sqrt{s}=$ 250, 300 and 500 GeV. Such a differential cross section gets significantly enhanced near the forward direction $\cos\theta \approx 1$, which is due to the $t$-channel diagram $(v_1)$ in Fig.\ref{fig:diag-ee2gah} as explained below. 

Finally, Fig.\ref{fig:fig6} displays differential cross section for $e^-e^+ \to \gamma h^0$ in both SM and HTM for two values of $\sqrt{s}=$ 250 GeV (left) and 500 GeV. As expected, in the absence of new Lorentz structure in the HTM, the differential cross section in the HTM possess the same shape as in the SM and is slightly shifted up due to the charged Higgses effects.

\section{Conclusions}
\label{conclu}
The LC is expected to play a crucial role in understanding the nature of the Higgs boson, 
which is just getting started, and will have a lot to add to whatever the LHC will find out. 
In this paper, we have studied, in the framework of the HTM, the one-loop processes 
$e^+e^-\to \gamma h^0$ and $e^-\gamma \to e^- h^0$ in the Feynman gauge using dimensional regularization for the future LC machine, where $h^0$ is the lightest, neutral, CP-even Higgs boson. 
We have shown that the singly (-doubly) charged Higgs loops in HTM can modify significantly the cross section compared to the SM predictions, depending on the parameter $\lambda_1$ and $\lambda_4$ which controls the contribution of the charged Higgs bosons in the loops. We find that such cross sections for the studied processes are quite sensitive to these parameters; so that the observable $R_{\gamma h^0}$ that we defined for the LC can  be away from unity implying the presence of new charged particles in the loops. Such new charged particles would also contribute to the one loop couplings $h\to \gamma \gamma$ and $h\to \gamma Z$. Therefore, we have shown that the correlation between $R_{\gamma h^0}$,  $R_{e^- h^0}$ and $R_{\gamma \gamma}(h^0)$ can be mainly positive for $\sqrt s$ = 250 GeV depending on the HTM parameter space. 
We also illustrate on one hand, the transverse momentum distribution for the $e^+e^-\to \gamma h^0$ which shows an enhancement near $p_T^{h^0}\approx \sqrt{s}/2$ and on the other hand the differential cross section
for $e^+e^-\to \gamma h^0$ for different center of mass energy.

\section*{Acknowledgment}

\noindent This work is supported in part by the Moroccan Ministry of Higher Education and Scientific Research under Contract N$^{\circ}$ PPR/2015/6, and by the GDRI-Physique de l'infiniment petit et l'infiniment grand-P2IM. AA acknowledges NCTS NTU for hospitality. LR would like to thank the L2C laboratory of the university of Montpellier for hospitality, where part of this work has been done.


\begin{thebibliography}{}

\bibitem{atlasdiscovery}
G.~Aad {\it et al.} [ATLAS Collaboration],
Phys.\ Lett.\ B {\bf 716} (2012) 1


\bibitem{cmsdiscovery}
S.~Chatrchyan {\it et al.} [CMS Collaboration],
Phys.\ Lett.\ B {\bf 716} (2012) 30

\bibitem{Ellis:2002wba}
J.~R.~Ellis,
hep-ph/0211168.
  
  
\bibitem{Weinberg:1979sa}
S.~Weinberg,
Phys.\ Rev.\ Lett.\  {\bf 43} (1979) 1566.



\bibitem{Wilczek:1979hc}
F.~Wilczek and A.~Zee,
Phys.\ Rev.\ Lett.\  {\bf 43} (1979) 1571.



\bibitem{Minkowski:1977sc}
P.~Minkowski,
Phys.\ Lett.\  {\bf 67B} (1977) 421.


\bibitem{Mohapatra:1979ia}
R.~N.~Mohapatra and G.~Senjanovic,
Phys.\ Rev.\ Lett.\  {\bf 44} (1980) 912.



\bibitem{Yanagida:1979as}
T.~Yanagida,
Conf.\ Proc.\ C {\bf 7902131} (1979) 95.


\bibitem{GellMann:1980vs}
M.~Gell-Mann, P.~Ramond and R.~Slansky,
Conf.\ Proc.\ C {\bf 790927} (1979) 315


\bibitem{Schechter:1980gr}
J.~Schechter and J.~W.~F.~Valle,
Phys.\ Rev.\ D {\bf 22} (1980) 2227.


\bibitem{Glashow:1979nm}
S.~L.~Glashow,
NATO Sci.\ Ser.\ B {\bf 61} (1980) 687.

\bibitem{Babu:1993qv}
K.~S.~Babu, C.~N.~Leung and J.~T.~Pantaleone,
Phys.\ Lett.\ B {\bf 319} (1993) 191



\bibitem{Antusch:2001vn}
S.~Antusch, M.~Drees, J.~Kersten, M.~Lindner and M.~Ratz,
Phys.\ Lett.\ B {\bf 525} (2002) 130



\bibitem{Magg:1980ut}
M.~Magg and C.~Wetterich,
Phys.\ Lett.\  {\bf 94B} (1980) 61.



\bibitem{Cheng:1980qt}
T.~P.~Cheng and L.~F.~Li,
Phys.\ Rev.\ D {\bf 22} (1980) 2860.


\bibitem{Lazarides:1980nt}
G.~Lazarides, Q.~Shafi and C.~Wetterich,
Nucl.\ Phys.\ B {\bf 181} (1981) 287.



\bibitem{Mohapatra:1980yp}
R.~N.~Mohapatra and G.~Senjanovic,
Phys.\ Rev.\ D {\bf 23} (1981) 165.



\bibitem{Foot:1988aq}
R.~Foot, H.~Lew, X.~G.~He and G.~C.~Joshi,
Z.\ Phys.\ C {\bf 44} (1989) 441.


\bibitem{Arhrib:2011uy}
A.~Arhrib, R.~Benbrik, M.~Chabab, G.~Moultaka, M.~C.~Peyranere, L.~Rahili and J.~Ramadan,
Phys.\ Rev.\ D {\bf 84} (2011) 095005
;
A.~Arhrib, R.~Benbrik, G.~Moultaka and L.~Rahili,
arXiv:1411.5645 [hep-ph].


\bibitem{Dev:2013ff}
P.~S.~Bhupal Dev, D.~K.~Ghosh, N.~Okada and I.~Saha,
JHEP {\bf 1303} (2013) 150
Erratum: [JHEP {\bf 1305} (2013) 049]
;
Y.~Du, A.~Dunbrack, M.~J.~Ramsey-Musolf and J.~H.~Yu,
JHEP {\bf 1901} (2019) 101

\bibitem{Perez:2008ha}
P.~Fileviez Perez, T.~Han, G.~y.~Huang, T.~Li and K.~Wang,
Phys.\ Rev.\ D {\bf 78} (2008) 015018



\bibitem{Melfo:2011nx}
A.~Melfo, M.~Nemevsek, F.~Nesti, G.~Senjanovic and Y.~Zhang,
Phys.\ Rev.\ D {\bf 85} (2012) 055018

\bibitem{delAguila:2008cj}
F.~del Aguila and J.~A.~Aguilar-Saavedra,
Nucl.\ Phys.\ B {\bf 813} (2009) 22



\bibitem{Chakrabarti:1998qy}
S.~Chakrabarti, D.~Choudhury, R.~M.~Godbole and B.~Mukhopadhyaya,
Phys.\ Lett.\ B {\bf 434} (1998) 347


\bibitem{Aoki:2011pz}
M.~Aoki, S.~Kanemura and K.~Yagyu,
Phys.\ Rev.\ D {\bf 85} (2012) 055007



\bibitem{Chun:2013vma}
E.~J.~Chun and P.~Sharma,
Phys.\ Lett.\ B {\bf 728} (2014) 256


\bibitem{Banerjee:2013hxa}
S.~Banerjee, M.~Frank and S.~K.~Rai,
Phys.\ Rev.\ D {\bf 89} (2014) no.7,  075005


\bibitem{Gianotti:2000tz} 
F.~Gianotti and M.~Pepe-Altarelli,
Nucl.\ Phys.\ Proc.\ Suppl.\  {\bf 89} (2000) 177,
D.~Zeppenfeld, R.~Kinnunen, A.~Nikitenko and E.~Richter-Was,
Phys.\ Rev.\ D {\bf 62} (2000) 013009


\bibitem{Moortgat-Picka:2015yla} 
G.~Moortgat-Pick {\it et al.},
Eur.\ Phys.\ J.\ C {\bf 75} (2015) no.8,  371
,  
H.~Baer {\it et al.},
arXiv:1306.6352 [hep-ph].


\bibitem{Barroso:1985et} 
A.~Barroso, J.~Pulido and J.~C.~Romao,
Nucl.\ Phys.\ B {\bf 267} (1986) 509.


\bibitem{Abbasabadi:1995rc} 
A.~Abbasabadi, D.~Bowser-Chao, D.~A.~Dicus and W.~W.~Repko,
Phys.\ Rev.\ D {\bf 52} (1995) 3919


\bibitem{Djouadi:1996ws} 
 A.~Djouadi, V.~Driesen, W.~Hollik and J.~Rosiek,
Nucl.\ Phys.\ B {\bf 491} (1997) 68


\bibitem{Akeroyd:1999gu} 
M.~Krawczyk, J.~Zochowski and P.~Mattig,
Eur.\ Phys.\ J.\ C {\bf 8} (1999) 495
;
A.~G.~Akeroyd, A.~Arhrib and M.~Capdequi Peyranere,
Mod.\ Phys.\ Lett.\ A {\bf 14} (1999) 2093
Erratum: [Mod.\ Phys.\ Lett.\ A {\bf 17} (2002) 373]


\bibitem{Kanemura:2018esc} 
S.~Kanemura, K.~Mawatari and K.~Sakurai,
Phys.\ Rev.\ D {\bf 99} (2019) no.3,  035023



\bibitem{Arhrib:2015wrb}
A.~Arhrib, R.~Benbrik and T.~C.~Yuan,
Eur.\ Phys.\ J.\ C {\bf 74} (2014) 2892


\bibitem{Demirci:2019ush}
M.~Demirci,
arXiv:1905.09363 [hep-ph].



\bibitem{bonilla2015}
C.~Bonilla, R.~M.~Fonseca and J.~W.~F.~Valle,
Phys. Rev. D {\bf 92} (2015) no.7,  075028



\bibitem{Mitra:2016wpr}
M.~Mitra, S.~Niyogi and M.~Spannowsky,
Phys.\ Rev.\ D {\bf 95} (2017) no.3,  035042

\bibitem{Agrawal:2018pci}
P.~Agrawal, M.~Mitra, S.~Niyogi, S.~Shil and M.~Spannowsky,
Phys.\ Rev.\ D {\bf 98} (2018) no.1,  015024
  
  

\bibitem{Abdallah:2002qj}
J.~Abdallah {\it et al.} [DELPHI Collaboration],
Phys.\ Lett.\ B {\bf 552} (2003) 127



\bibitem{CMS-PAS-HIG-16-036}
CMS Collaboration [CMS Collaboration],
``A search for doubly-charged Higgs boson production in three and four lepton final states at $\sqrt{s}=13~\mathrm{TeV}$,''
CMS-PAS-HIG-16-036.




\bibitem{Aaboud:2017qph}
M.~Aaboud {\it et al.} [ATLAS Collaboration],
Eur.\ Phys.\ J.\ C {\bf 78} (2018) no.3,  199



\bibitem{Kanemura:2013vxa}
S.~Kanemura, K.~Yagyu and H.~Yokoya,
Phys.\ Lett.\ B {\bf 726} (2013) 316


\bibitem{Kanemura:2014goa}
S.~Kanemura, M.~Kikuchi, K.~Yagyu and H.~Yokoya, 
Phys.\ Rev.\ D {\bf 90} (2014) 115018.



\bibitem{Kanemura:2014ipa}
S.~Kanemura, M.~Kikuchi, H.~Yokoya and K.~Yagyu,
PTEP {\bf 2015} (2015) 051B02


\bibitem{Aaboud:2018qcu}
M.~Aaboud {\it et al.} [ATLAS Collaboration],
Eur.\ Phys.\ J.\ C {\bf 79} (2019) no.1,  58


\bibitem{Khachatryan:2014sta}
V.~Khachatryan {\it et al.} [CMS Collaboration],
Phys.\ Rev.\ Lett.\  {\bf 114} (2015) no.5,  051801


\bibitem{Sirunyan:2017ret}
A.~M.~Sirunyan {\it et al.} [CMS Collaboration],
Phys.\ Rev.\ Lett.\  {\bf 120} (2018) no.8,  081801




\bibitem{Ghosh:2017pxl}
D.~K.~Ghosh, N.~Ghosh, I.~Saha and A.~Shaw,
Phys.\ Rev.\ D {\bf 97} (2018) no.11,  115022


\bibitem{Bechtle2014:HB}
P.~Bechtle, O.~Brein, S.~Heinemeyer, O.~St{\aa}l, T.~Stefaniak, G.~Weiglein and K.~E.~Williams,
Eur.\ Phys.\ J.\ C {\bf 74} (2014) no.3,  2693
 
 
\bibitem{Bechtle2014:HS}
P.~Bechtle, S.~Heinemeyer, O.~St{\aa}l, T.~Stefaniak and G.~Weiglein,
Eur.\ Phys.\ J.\ C {\bf 74} (2014) no.2,  2711



\bibitem{FA2}
T.~Hahn,
Comput.\ Phys.\ Commun.\  {\bf 140} (2001) 418
;
T.~Hahn and C.~Schappacher,
Comput.\ Phys.\ Commun.\  {\bf 143} (2002) 54
;
T.~Hahn and M.~Perez-Victoria,
Comput.\ Phys.\ Commun.\  {\bf 118} (1999) 153
;
J.~Kublbeck, M.~Bohm and A.~Denner,
Comput.\ Phys.\ Commun.\  {\bf 60} (1990) 165.
.

\bibitem{FF} 
G.~J.~van Oldenborgh,
Comput.\ Phys.\ Commun.\  {\bf 66} (1991) 1.
;
T.~Hahn,
Acta Phys.\ Polon.\ B {\bf 30} (1999) 3469
;
T.~Hahn,
PoS ACAT {\bf 2010} (2010) 078.

\bibitem{chabab2016}
M.~Chabab, M.~C.~Peyranère and L.~Rahili,
Phys.\ Rev.\ D {\bf 93} (2016) no.11,  115021


\bibitem{Aaboud:2018xdt}
M.~Aaboud {\it et al.} [ATLAS Collaboration],
Phys.\ Rev.\ D {\bf 98} (2018) 052005

\bibitem{Sirunyan:2018ouh}
A.~M.~Sirunyan {\it et al.} [CMS Collaboration],
JHEP {\bf 1811} (2018) 185
  
\end{thebibliography}
\end{document}